\documentclass{aa} 
\usepackage[utf8]{inputenc}
\usepackage{appendix}
\usepackage{enumitem} 
\usepackage{graphicx}
\usepackage{txfonts}
\usepackage{natbib}
\usepackage{amssymb}%http://ctan.org/pkg/amssymb
\usepackage{pifont}%http://ctan.org/pkg/pifont
\usepackage{subcaption}
\def\msun{{\rm\,M_\odot}}
\def\vmi{\hbox{\it V--I\/}}
\def\bmv{\hbox{\it B--V\/}}
\def\bmi{\hbox{\it B--I\/}}
\def\hst{{\it HST\/}}
\newcommand{\cmark}{\ding{51}}%
\bibpunct{(}{)}{;}{a}{}{,} % to follow the A&A style
\begin{document} 

   \title{Horizontal branch morphology: A new photometric parametrization}
   \author{M.~Torelli \inst{\ref{inst1},} \inst{\ref{inst2}}\and  G.~Iannicola  \inst{\ref{inst2}} \and P.B.~Stetson\inst{\ref{inst3}}\and I.~Ferraro\inst{\ref{inst2}} \and G.~Bono\inst{\ref{inst1},}\inst{\ref{inst2}} \and M.~Salaris\inst{\ref{inst4}}  \and M.~Castellani\inst{\ref{inst2}} \and M.~Dall'Ora\inst{\ref{inst5}} \and A.~Fontana\inst{\ref{inst2}} \and M.~Monelli\inst{\ref{inst6},} \inst{\ref{inst7}}\and A.~Pietrinferni\inst{\ref{inst8}}   
             }
   \institute{Department of Physics, Universit\`{a} di Roma Tor Vergata, via della Ricerca Scientifica 1, I-00133 Roma, Italy \label{inst1}, \email{marianna.torelli@inaf.it}
     \and INAF-Osservatorio Astronomico di Roma,via Frascati 33, I-00078 Monte Porzio Catone, Italy \label{inst2}
     \and NRC-Herzberg, Dominion Astrophysical Observatory,  5071 West Saanich Road, Victoria BC V9E 2E7, Canada \label{inst3}
     \and Astrophysics  Research  Institute,  Liverpool  John  Moores  University,  IC2  Building,  Liverpool  Science  Park,  146  Brownlow  Hill,Liverpool L3 5RF, UK \label{inst4}
     \and INAF-Osservatorio Astronomico di Capodimonte, Salita Moiariello 16, I-80131 Napoli, Italy \label{inst5}
     \and IAC - Instituto de Astrofisica de Canarias, Calle Via Lactea s/n, E38205 La Laguna, Spain \label{inst6}
     \and Departamento de Astrofisica, Universidad de La Laguna, Avenida Astrofísico Francisco Sánchez s/n, E38200 Tenerife, Spain\label{inst7}
     \and INAF - Osservatorio Astronomico d'Abruzzo, Via Mentore Maggini snc, Loc. Collurania, I-64100 Teramo, Italy \label{inst8}
     }
                     
\date{Received XXX; accepted XXX}
% \abstract{}{}{}{}{} 
% 5 {} token are mandatory
  \abstract 
 {Theory and observations indicate that the distribution of stars along the horizontal
   branch of Galactic globular clusters mainly depends on the metal content. 
   However, the existence of globular clusters with
   similar metal content and absolute age but different horizontal branch 
   morphologies, suggests the presence  of another parameter affecting
   the star distribution along the branch.
    }
   {To investigate the variation of the horizontal branch 
   morphology in Galactic globular clusters,
we define a new photometric horizontal branch morphology index, 
overcoming some of the limitations and degeneracies affecting similar 
indices available in the literature.
    }
   {We took advantage of a sample of 64 Galactic globular clusters, 
   with both space-based imaging data (\textit{Advanced Camera for Surveys
   survey of Galactic globular clusters}) and homogeneous ground-based photometric
   catalogues in five different bands ($U$,$B$,$V$, $R$, $I$).  
   The new index, $\tau_{HB}$, is defined as the ratio
   between the areas subtended by the cumulative number distribution in magnitude
   ($I$) and in colour ($\vmi$) of all stars along the horizontal branch.
   } 
   {This new index shows a linear trend over the entire range in metallicity
    (-2.35 $\leq$ [Fe/H] $\leq$ -0.12) covered by our Galactic 
    globular cluster sample. We found a linear relation between $\tau_{HB}$ and absolute
   cluster ages. We also found a quadratic anti-correlation with [Fe/H], becoming linear when we eliminate the age effect on $\tau_{HB}$ values. 
  Moreover, we identified a subsample of eight clusters that are peculiar according to their $\tau_{HB}$ values. 
  These clusters have bluer horizontal branch morphology when compared to typical ones of similar metallicity. 
  These findings allow us to define them as the 'second parameter' clusters in the sample. 
A comparison with synthetic horizontal branch models suggests that they
cannot be entirely explained with a spread in helium content.}
{}

   \maketitle
%
%_______________________________________________________________________________
\section{Introduction}\label{sec:1}
Globular clusters are stellar systems that play a fundamental role in constraining 
the formation and evolution of galaxies \citep{searlezinn,vand,leaman} and cosmological 
parameters. Dating back more than half a century \citep{sandage}, the 
absolute ages of Galactic globular clusters (GGCs) have been used to provide a lower limit to 
the age of the Universe 
\citep[see e.g.][and references therein]{salarisweiss, dotter11, monelli,richer}  
and estimate of the primordial helium abundance \citep[see e.g.][]{zoccali, salarisbook,villanova}. GGCs are also 
laboratories to investigate evolutionary \citep{chaboyer, dotter07, brown, weiss} 
and pulsational \citep{bono94, bono99,marconi}  
properties of old, low-mass stars.
In this context, advanced evolutionary phases (red giant [RGs] and horizontal
branch [HB] stars) have several advantages when compared with 
main sequence (MS) stars. They are a few magnitudes brighter and 
within the reach of spectroscopic investigations at the 8-10m class 
telescopes. This means that they can be used for investigating the cluster 
dynamical evolution and interaction with the Galactic potential 
\citep{pancino, zocchi, calamida,lanzoni}. 
Moreover, their chemical abundances (iron peak, $\alpha$-, 
neutron capture elements) can be measured with high accuracy both in 
the optical \citep{carretta14} and in the near-infrared (NIR) regime \citep{dorazi}. 

Despite a general consistency between theory and observations 
concerning hydrogen and helium burning phases, 
we continue
to face a number of long-standing open questions. Amongst them 
the morphology of the HB 
plays a pivotal role. Stars along the HB are low-mass 
(M $\approx$ 0.50-0.80 $\msun$), core-helium-burning stars and their 
distribution along the HB depends, at fixed initial chemical composition, 
on their  
envelope mass. Indeed, the envelope mass when moving from the red HB 
(RHB) to the extreme HB (EHB) decreases from $M_{env} \approx$0.30 $\msun$
to $M_{env} \approx$0.0001 $\msun$, while the effective temperature increases 
from $\approx$ 5,500 $\mathrm{K}$ to $\approx$30,000 $\mathrm{K}$. 

The current empirical and theoretical evidence indicates that the HB 
morphology is affected by the initial metal content. Metal-poor clusters are mainly 
characterized by a blue HB morphology. This means that in these clusters HB 
stars are mainly distributed along the blue, hot, and extremely hot region.    
Metal-rich GGCs are generally characterized by a red HB morphology, that is, HB stars 
in these clusters are mainly distributed in the red (cool) region. 

Although this theoretical and empirical framework appears well established, 
there is solid evidence that GGCs with similar chemical compositions display 
different HB morphologies. This suggests that 
 the HB morphology was  affected by at least a 'second parameter' 
\citep[see. e.g.][and references therein]{sandagewildey}.  
This problem was defined as the 'second parameter problem' and the clusters affected 
by this problem were called second parameter clusters.

During the last half-century several working hypotheses have been suggested 
to explain the second parameter problem. They include variations of the initial 
helium content \citep{vandenbergh,sandagewildey},  
dynamical effects related to the cluster mass \citep[e.g.][]{recioblanco}, 
the cluster age \citep[e.g.][]{searlezinn},
dynamical evolution \citep[e.g.][]{iannicola}, 
or a combination of two or more of these -- for example age and/or metallicity 
plus helium content, as suggested by \citet{freemannorris,gratton}. 

To quantify the extent of this second parameter problem,
\citet{leethesis,lee2} suggested an HB morphology 
index based on star counts along the HB. It is defined as 
the difference between the number of stars that are bluer (B) and redder (R) than the 
RR Lyrae (RRL) instability strip, divided by the sum of the number of 
blue, red, and variable (V) stars: 
$HBR=\frac{B-R}{B+R+V}$. This HB morphology index has been quite popular, 
because it can be easily estimated from the theoretical 
and the observational point of view (star counts). 
However, it is prone to intrinsic degeneracies in both the metal-poor 
and the metal-rich regimes, in the sense that when the observed HB 
is bluer or redder than the RRL strip, $HBR$ stays constant, 
irrespectively of the exact distribution of stars along the HB.
 
Similar HB morphology indices have been suggested in the literature, but 
using different cuts in colour. In particular, \citet{buonanno} suggested 
splitting HB stars bluer than the RRL instability strip into two blocks: 
stars hotter than the RRL instability strip and cooler than 
($B$-$V$)=-0.02 were called B1, while those hotter than ($B$-$V$)=-0.02 were 
called B2. The new index had the key advantage of removing the degeneracy 
of the $HBR$ index in the metal-poor regime, but it was still affected by 
degeneracies in the metal-rich regime. 

Considering that almost all GGCs show the presence of 
multiple populations, \citet{milone} introduced two new indices for 
describing the HB morphology. L1 is the difference in colour between 
the red giant branch (RGB) and the coolest red point of the HB, while L2 is 
the colour extension of the HB. These two indices allowed the authors to identify 
three different GGC groups from the L1-[Fe/H] diagram, and to find correlations
of L1 with cluster age and metallicity. In addition, they found a variation of 
L2 with cluster luminosity (mass) and with helium content. This latter 
correlation is connected with the presence of multiple populations in 
globular clusters and it could be an additional ingredient to explain the 
HB colour extension. 
These two parameters are interesting, because they are correlated with 
the physics characterizing the HB stars. However, their definition seems 
to be very sensitive to the choice of the key points selected in colour magnitude
diagrams (HB luminosity level, RGB).

In this work, we introduce a new HB morphology index based on 
the ratio between the areas subtended by the cumulative number distribution (CND)
of star counts along the observed cluster HB in magnitude 
($I$-band) and in colour ($\vmi$): $\tau_{HB}= A_{CND}(I)/A_{CND}(\vmi)$. 
This new index has been calculated in a large sample (64) 
of GGCs, for which both space- and ground-based optical  
photometric catalogues are available. For the same sample of GGCs, we have also estimated the classical $HBR$ index and performed a detailed comparison with 
$\tau_{HB}$. We present evidence that our new index, in contrast with 
similar indices available in the literature, shows a well-defined correlation
with cluster age and an anti-correlation with cluster iron abundance. 

The structure of the paper is the following. 
We present in Sect.~\ref{sec:2} our sample of GGCs with their 
photometric coverage. We also describe 
the method adopted for the separation between candidate cluster and 
field stars based on the comparison of star spectral energy distributions 
(SEDs).  Section~\ref{sec:3} deals with the classical HB morphology index 
$HBR$. We analyse its pros and cons and 
calculate its values for the entire cluster sample, together 
with its dependence on age and metallicity, while in 
Sect.~\ref{sec:4} we do the same analysis for the L1 and L2 indices.  
In Sect.~\ref{sec:5} we introduce our 
new HB morphology index, $\tau_{HB}$, calculate its values for 
our GGC sample, and discuss pros and cons 
compared to $HBR$. In
Sect.~\ref{sec:6} we analyse and compare the relative difference between
the classical HB morphology index and $\tau_{HB}$ estimates 
when considering just space- or just ground-based observations. The correlation 
of our new index with the absolute age and the metallicity of the individual 
clusters is addressed in Sect.~\ref{sec:7}. Here we also identify the 
{second parameter} clusters in our sample that show very different 
estimates in $\tau_{HB}$ compared to the ones attained by globulars with similar [Fe/H] values.
In Sect.~\ref{sec:8} we compare  $\tau_{HB}$  
with synthetic HB models specifically 
computed for this work. The summary of the results and a brief discussion concerning 
future developments of the project are outlined in Sect.~\ref{sec:9}. 
In Appendix~\ref{app:A} we list specific information on individual 
GGCs for which the photometric properties are uncertain either due to the 
lack of ground-based or space-based data, because the photometry does not 
cover the entire cluster area (tidal radius), or because of the small 
number of HB stars. In Appendix~\ref{app:B} we also included a few notes for the GGCs we 
defined as 'outliers' in the analysis of the HB morphology.

%
%_______________________________________________________________________________
\section{Globular cluster sample}\label{sec:2}

In this work we use a sample of 64 GGCs for which optical images are 
available from both space-based (Advanced Camera for Surveys (ACS)/Wide
Field Channel (WFC) on board the
Hubble Space Telescope ($\hst$)) and ground-based observations. 
The relevant parameters for our cluster sample are listed in Table~\ref{table:gcsuni}.

\begin{figure}
\centering
\includegraphics[width=\hsize]{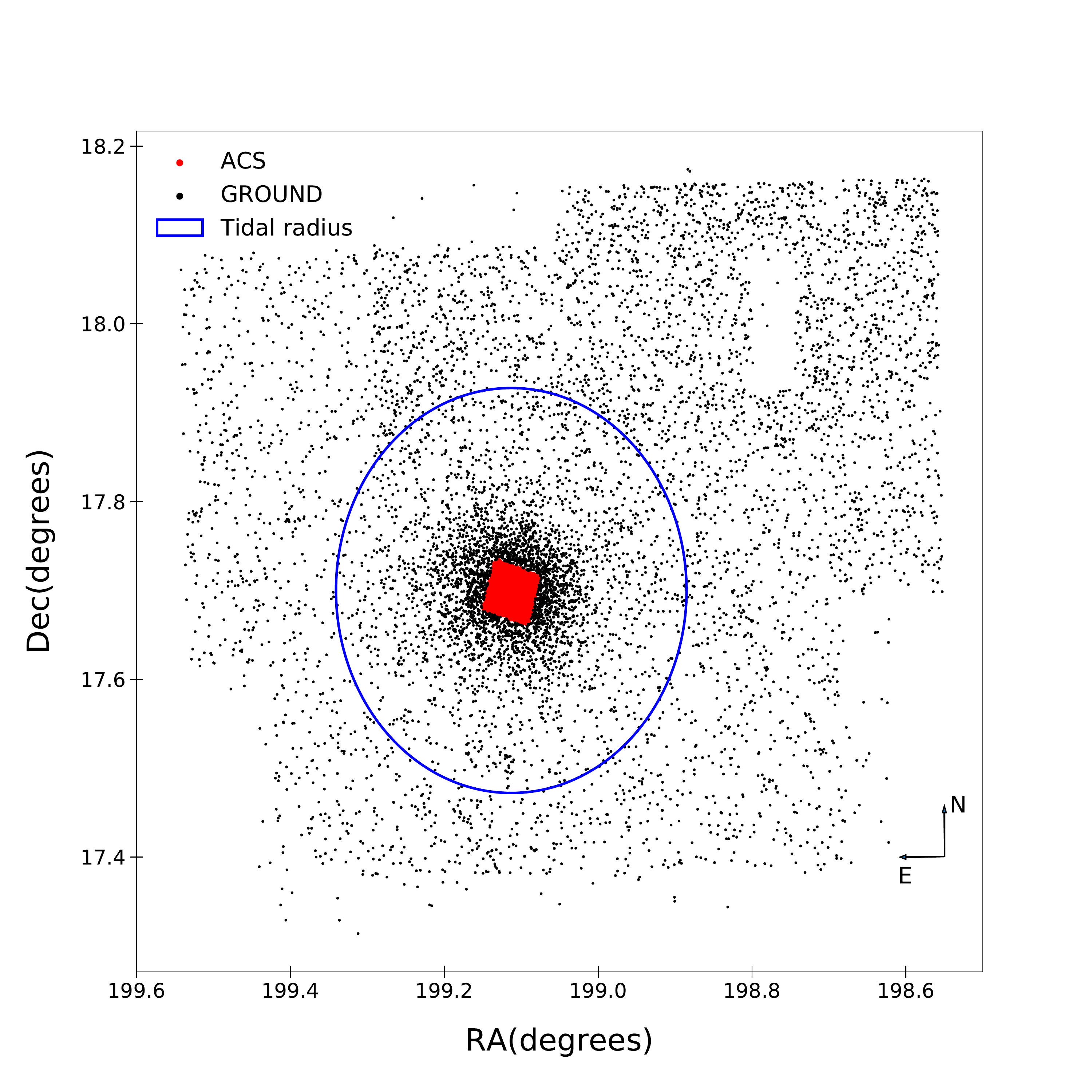}
\caption{Sky distribution of ground-based (black dots) and space-based (red dots) data for the cluster NGC~5053. North is up and east is to the left.}
\label{geom}
\end{figure}

\begin{figure}
\centering
\includegraphics[width=\hsize]{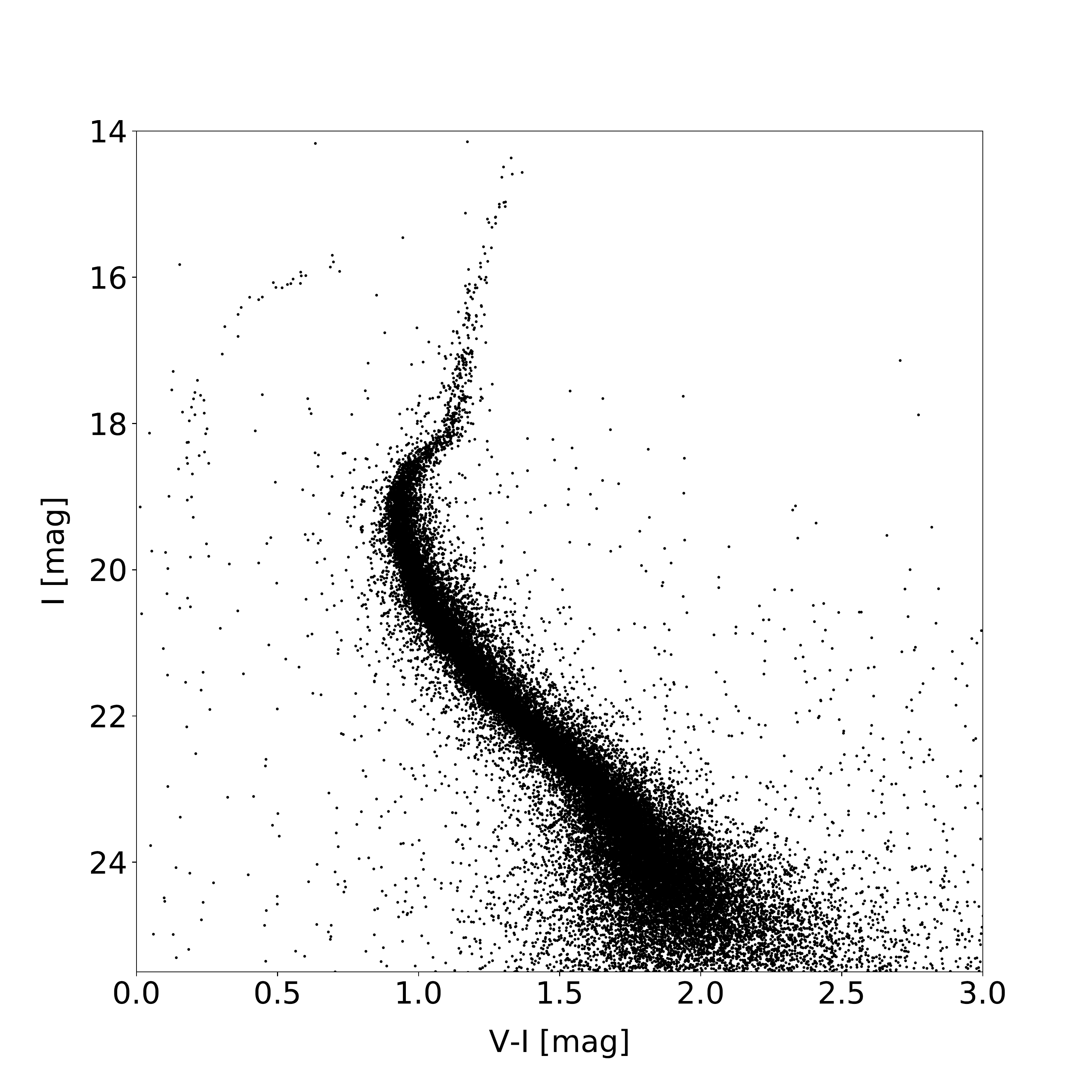}
\caption{Cluster NGC~5286 $I$, $\vmi$ CMD based only on ACS-$\hst$ data.}
\label{CMDACS}
\end{figure}

We took advantage of the photometric catalogues provided by \citet{sara,dotter11} in
the context of the Hubble Space Telescope Treasury project, \textit{An ACS Survey
of Galactic Globular Clusters}
\footnote{Photometry available at 
\url{https://www.astro.ufl.edu/~ata/public_hstgc/}}.
This survey is based on a single ACS pointing across the centre of each cluster,
observed through two complete orbits, one for the $F606W$ ($\sim V$)
and one for the $F814W$ ($\sim I$). Cluster NGC 6715 represents an exception since it was observed for two orbits in each filter \citep{anderson}.
Thanks to these data we can avoid the crowding problems in the central regions of the
clusters (the red region in Fig.~\ref{geom} shows the field coverage for one of the
globulars in the sample, \object{NGC 5053}) and reach very faint magnitudes ($I \sim 26.0$, 
see ACS colour-magnitude diagram (CMD) in Fig.~\ref{CMDACS}).

To cover a significant fraction of the body of each cluster  
we adopted the multi-band ($U$, $B$, $V$, $R$, $I$) optical catalogues provided by one of the authors 
\citep[PBS, see for example][]{stetson05,stetson14,stetson19}, 
based on images collected with several ground-based telescopes. The ground-based data allow us to reach the tidal radius for most of the globulars. 
The blue line in Fig.~\ref{geom} shows the tidal radius of NGC~5053. There are a 
few clusters (see Appendix~\ref{app:A}) for which we cannot reach the tidal radius with 
our data while still covering the main body of the cluster. 
For 47 GGCs we have data in all available photometric bands from both space- 
and ground-based facilities. 

Thanks to the fact that all the clusters in the sample have data in $B$, $V$,
$I$ bands (except for \object{NGC 6426}, \object{NGC 6624}, \object{NGC 6652}, 
\object{Lynga 7}, and \object{Palomar 2}, see Appendix~\ref{app:A}),
we based the cluster and field stars' separation on the SED of the stars in these
bands, following the procedure described in \citet{dicecco,calamida}.
We selected first the {bona fide} cluster stars considering only the central region, 
since we expect that it is less contaminated by field stars. Then we compared the SEDs of these 
stars to the ones in the total catalogue, to separate cluster stars from the field.
We note that for stars with space data only we did not make any selection, 
because we expect negligible contamination by field stars in the ACS field of 
view (FoV). 
Therefore, in this case we took advantage of the entire ACS catalogues.

Moreover, when we had measurements in the $V$ and $I$ bands from both ACS and ground-based telescopes, 
we preferred the first ones for their higher signal-to-noise ratio (S/N) values. We used the ACS coverage for the central region 
of our globulars within the ACS FoV ($\sim 4 \arcmin $), while outside up to the tidal radius we adopted the ground-based 
observations.

\begin{figure}
\centering
\includegraphics[width=\hsize]{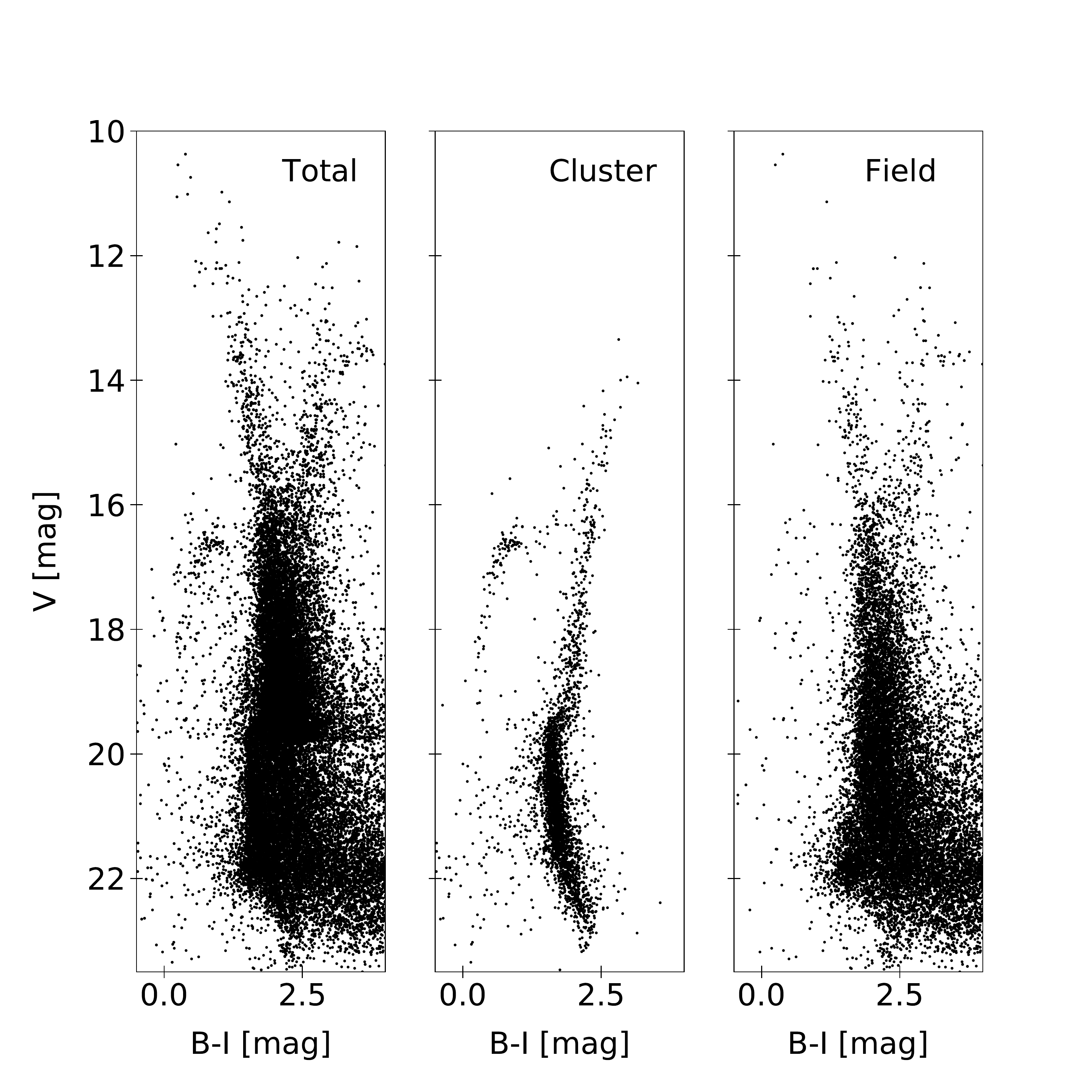}
\caption{Cluster NGC~5286 $V$, $\bmi$ CMD. The left panel shows all stars. The central panel shows our 
candidate cluster stars. The right panel shows the 
candidate field stars. Candidate cluster and field stars have been selected following the method described in the main text.}
\label{CMD}
\end{figure}

An example of our selection is shown in Fig.~\ref{CMD}, which displays the $V$, $\bmi$ CMD of \object{NGC 5286} for total (left panel), 
cluster (central panel), and field (right panel) stars. We can appreciate from the figure that the joint catalogues 
allow us to cover the entire evolution of the stars in the globulars, from the faint part of the MS 
to the brighter region of the RGB-tip and asymptotic giant branch (AGB) in the CMD.
%FIGURE 1 
%FIGURE 2 
%FIGURE 3 
%_______________________________________________________________________________
\section{The $HBR$ morphology index}\label{sec:3}
As already discussed in Sect.~\ref{sec:1}, 
the traditional HB morphology index is defined as

\begin{equation}{
HBR = \frac{B-R}{B+R+V}
} ,\end{equation}

\noindent where $B$ is the number of HB stars bluer than the blue (hot) edge 
of the RRL instability strip, $V$ is the number of RR Lyrae stars, 
and $R$ is the number of HB stars redder than the red (cold) edge of 
the RRL instability strip. This HB morphology index has several 
advantages when compared with other observables (HB luminosity level, colour distribution) 
connected with the helium burning phases. 

First, when considered with the comparison between 
predicted and observed HB luminosity levels, HB star counts are 
independent of uncertainties affecting the cluster distance and 
are minimally affected by uncertainties in cluster reddening and the 
possible occurrence of differential reddening. Moreover, the comparison between the predicted and observed number of HB stars
is less prone to uncertainties affecting the transformations of theoretical 
predictions into the observational plane.

In the comparison between theory and observations, the $HBR$ index 
accounts for the global distribution of stars along the HB. On the other 
hand, the comparison between the predicted and observed HB luminosity levels 
is mainly restricted to the RRL instability strip, that is, the truly 
horizontal region of the HB. However, a minor fraction of GGCs hosts a sizeable
sample of RRLs to properly define the HB luminosity level.

One of the main cons of the $HBR$ index is that it describes a global property of HB stars. This 
means that it does not trace the detailed stellar distribution along the HB. It is also affected by
severe degeneracies both in the metal-poor and in the metal-rich regimes. 
This means that a significant fraction of GGCs attain values close either to 1 (blue HB morphology) or 
to -1 (red HB morphology), irrespective of the exact colour distribution 
of stars. 
  
Despite the stated limitations we calculated values of the $HBR$ index 
for our GGC sample. For each cluster 
we defined two boxes including candidate blue (B) and red (R) 
HB stars. The colour ranges for the boxes were fixed using 
either predicted or empirical edges for the RRL instability strip. 
Stars located in the RRL instability strip were neglected. 
To account for the number of variables 
we took advantage of the Clement catalogue of variable stars in GGCs. 
This catalogue was originally presented in \citet{clement}, but it is constantly 
updated\footnote{\url{http://www.astro.utoronto.ca/~cclement/cat/listngc.html}}.
For each cluster, it gives the number of variable stars together with their main 
pulsation parameters: pulsation period, mean magnitude and colour, and luminosity 
amplitude. For our analysis we took into account only confirmed cluster RRL stars, meaning that candidate cluster RRLs were not included. 

Throughout our analysis, we adopted the homogeneous metallicity scale provided by  
\citet{carretta}. The authors estimated the iron abundances from
high resolution spectra collected with the Fibre Large Srray Multi Element 
Spectrograph (FLAMES) at the Very Large Telescope (VLT), covering the entire 
metallicity range of GGCs. To increase the sample size  they also re-calibrated 
the most common metallicity scales available in the literature 
\citep{zw84,ki03,cg97}.  
For two clusters in our sample (NGC~6624 and Lynga~7) we adopted the metallicity 
estimate available in the 2010 version of the \citet{harris} 
catalogue\footnote{\url{http://physwww.mcmaster.ca/~harris/mwgc.dat}}.

%%%%% FIGURE HBR %%%%%%%%%%%%%%%%%%%
% FIGURE 4   
\begin{figure}
\centering
\includegraphics[width=\hsize]{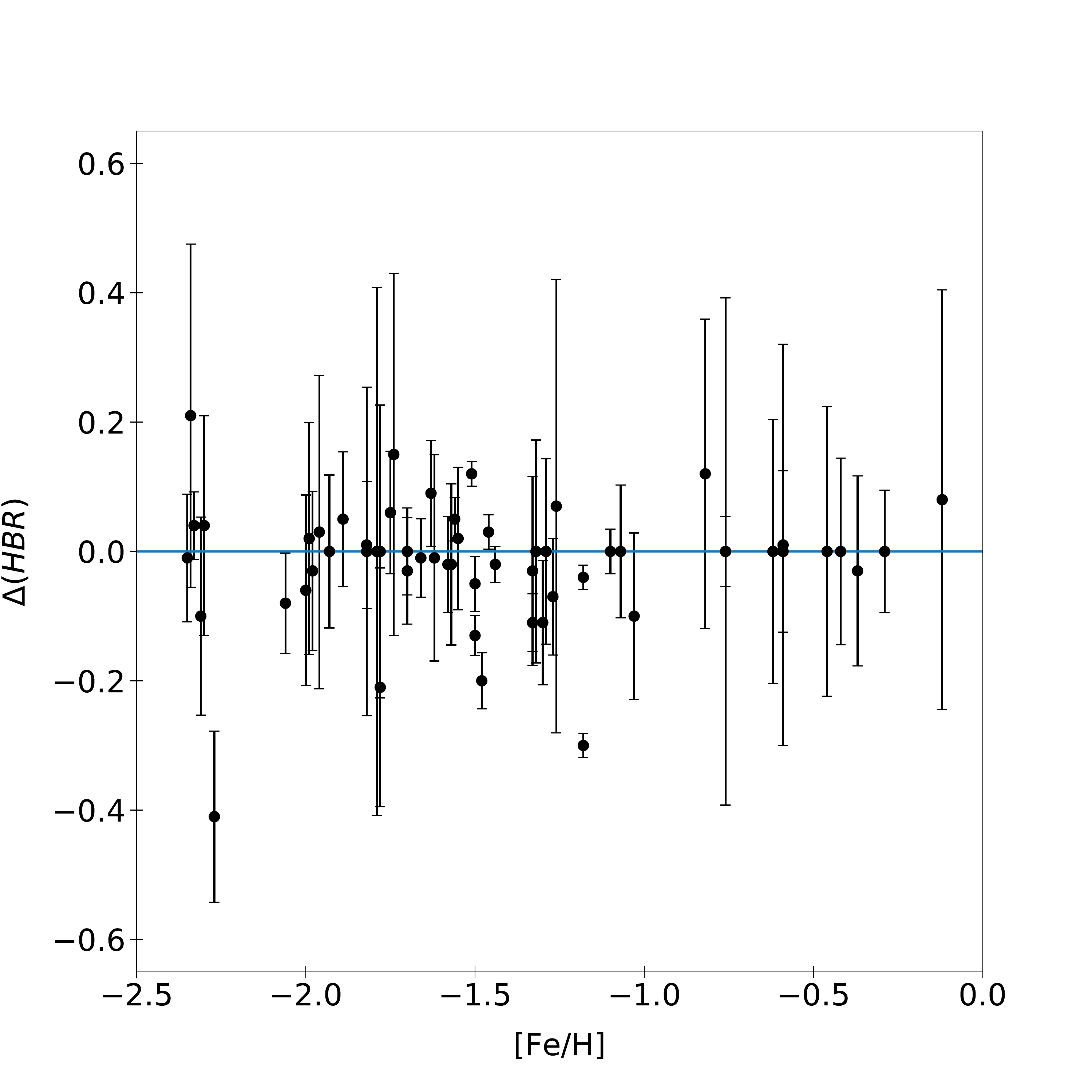} 
\caption{Difference between \citet{harris} values (2003 version) for $HBR$ and 
our measurements as a function of [Fe/H].}
\label{fig:DELTAHBRFEH}
\end{figure}

To validate the current estimates of the $HBR$ index with similar estimates 
available in the literature, Fig.~\ref{fig:DELTAHBRFEH} shows the comparison 
of our values with those provided by Harris in the 2003 version 
of his catalogue \citep{harris}. 
Data plotted in this figure show that both sets agree 
well over the entire metallicity range. Indeed, the difference is on average 
smaller than 20\%. There are a few outliers (\object{NGC 1851} and \object{NGC 4590}) for which the difference is 
approximately 30\%-40\%, but they are affected by a higher field-star contamination than the others,
despite our homogeneous cluster and field star separation.  

For each GGC in our sample, Table~\ref{table:hbr} gives its name 
(columns 1 and 2), the metallicity (column 3), and the number of blue (B) and red (R) stars (columns 4 and 5). 
In column 6 the values of the $HBR'$ index, defined as $HBR' = HBR+2$, are listed. 
The reason for this change is the following. From Fig.~\ref{fig:DELTAHBRFEH} we can see that the errors 
on the $HBR$ index are small and not realistic.
We define the uncertainty of the measured $HBR$ as 
\begin{equation}
\sigma(HBR)=\frac{B-R}{B+R+V}\cdot\Bigg(\frac{1}{\sqrt{B-R}}+\frac{1}{\sqrt{B+R+V}}\Bigg)
,\end{equation}
which takes into account the Poisson uncertainties in the star counts.
The errors provided by this formula vanish as soon as $B$ and $R$ attain 
similar values, that is when $HBR$ approaches zero. This is an intrinsic 
limitation affecting the definition of the $HBR$ index. To overcome this 
problem we decided to use $HBR'$. 

% FIGURE 5 
\begin{figure}
\centering
\includegraphics[width=\hsize]{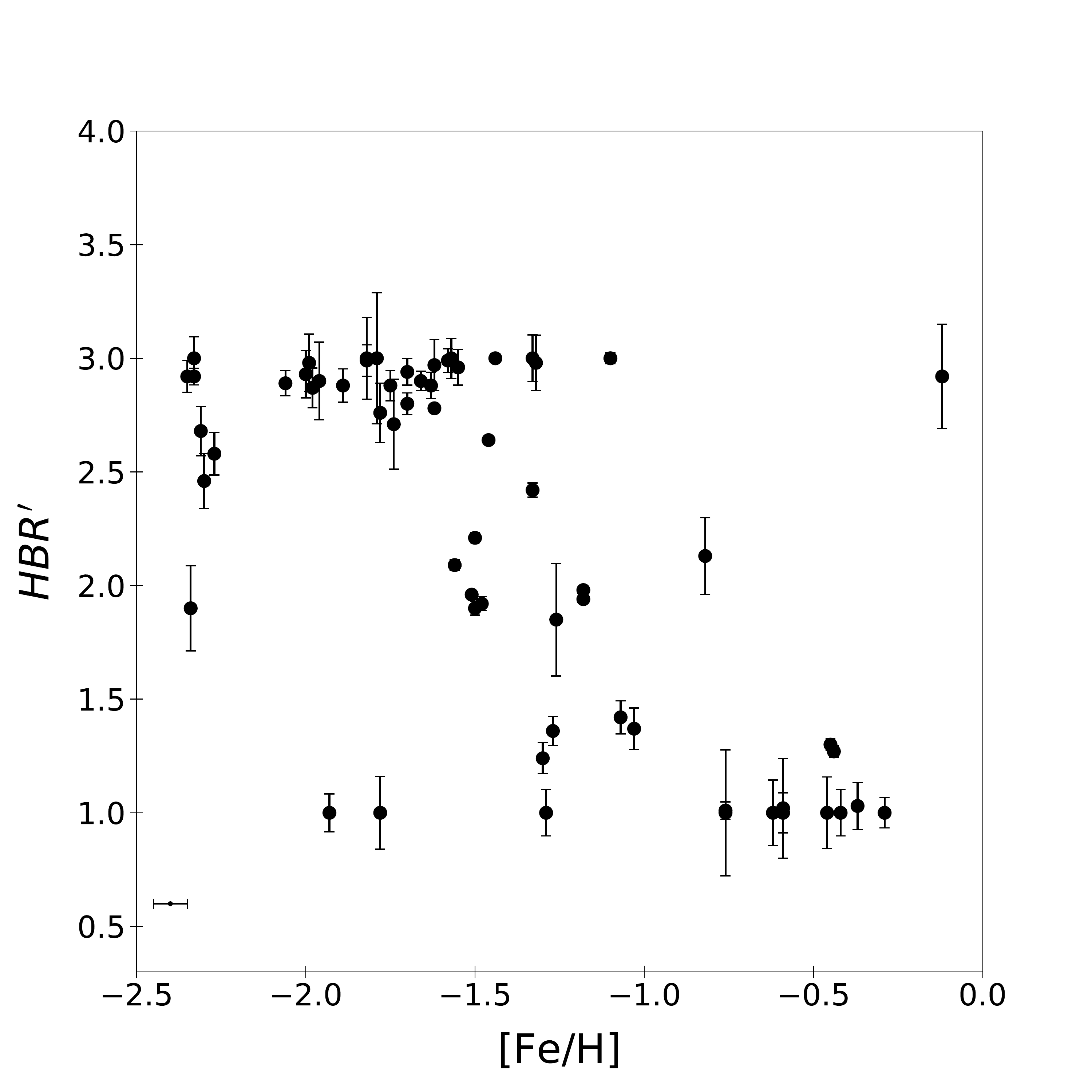} 
\caption{Observed $HBR'$ values as a function of metal content. The 
metallicity scale is from \citet{carretta} (see the Appendix for more details). 
The error bar in the lower left corner 
gives the uncertainty of the metal content (0.1 dex).}
\label{fig:HBRFEH}
\end{figure}
% FIGURE 6
\begin{figure}
\centering
\includegraphics[width=\hsize]{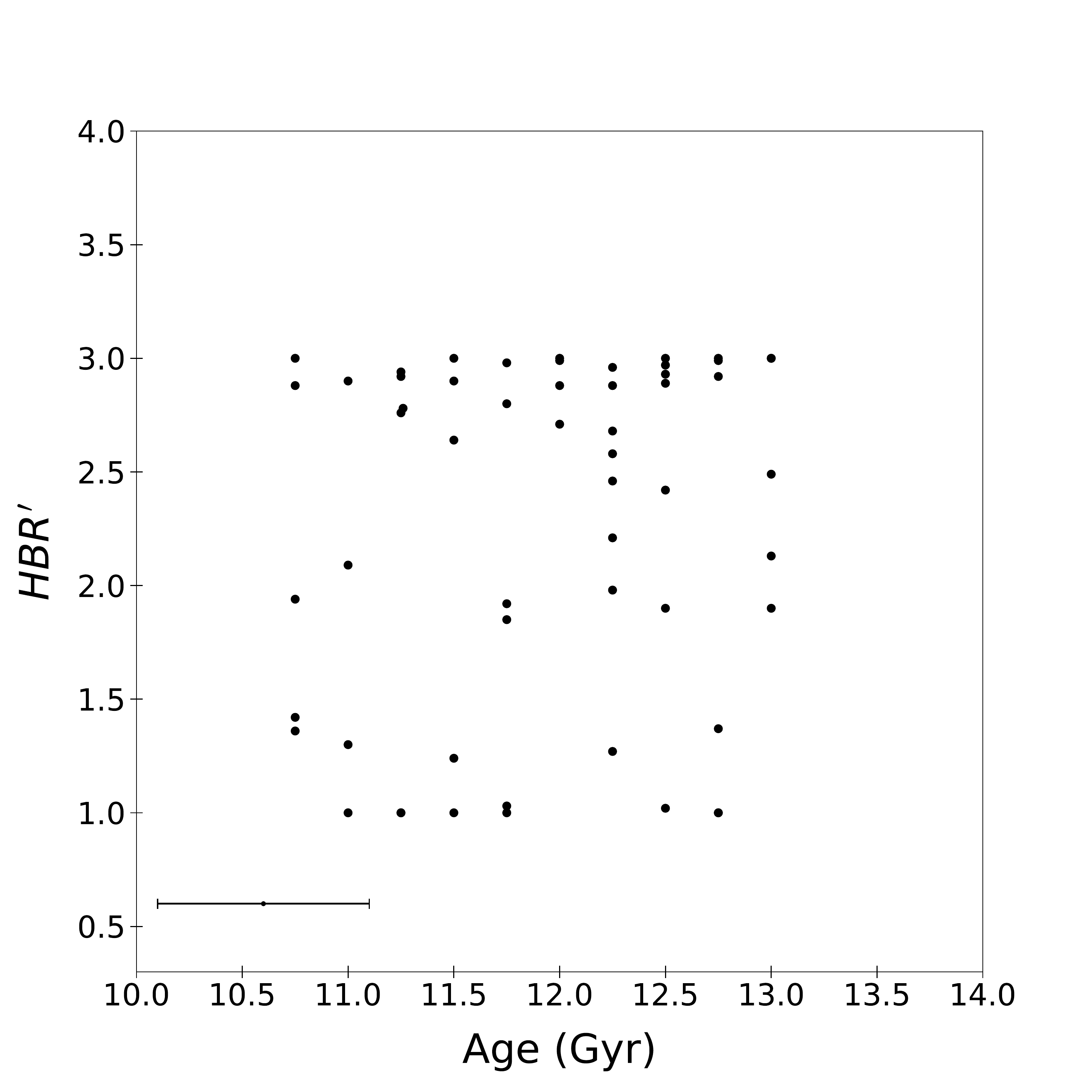} 
\caption{Observed $HBR'$ values as a function of the cluster 
ages (gigayears (Gyr)) provided by \citet{vand,leaman}. 
The error bar in the lower left corner gives the uncertainty in the cluster 
ages ($\pm$ 0.5~Gyr).}
\label{fig:HBRAGE}
\end{figure}

To investigate the dependence of the HB morphology on intrinsic 
cluster parameters, Fig.~\ref{fig:HBRFEH} shows the $HBR'$ index 
as a function of iron content for the current sample. 
Data plotted in this figure display some well established correlations. The HB morphology becomes systematically bluer when moving from 
metal-poor to more metal-rich GGCs.  

There is a well-defined degeneracy in the metal-intermediate regime 
(-1.6 $\le$ [Fe/H] $\le$\ -1)  in which GGCs with very similar metal abundances 
(the typical error on cluster metallicity being on 
the order of 0.1 dex -- see horizontal error bar in the left bottom 
corner) cover a broad range in $HBR'$ values. This means a variation 
from a very blue to a very red HB morphology, the 
well known second parameter problem.
Figure~\ref{fig:HBRFEH} shows that $HBR’$ is affected by degeneracy in its 
extreme values, $HBR' \sim -1$ (metal-rich regime) and $HBR'\sim +1$ (metal-poor regime). 
This means that  $HBR'$ has a low dynamical range attaining
similar values for clusters with different metallicity.

Cluster age has been considered together with iron 
content as one of the main culprits affecting the HB morphology.    
We took advantage of the homogeneous cluster age estimates 
provided by \citet{vand,leaman} 
for the same GGCs, to investigate the correlation with $HBR'$ index.
They estimated the cluster ages as follows.
Firstly, they de-reddened the HB stars and then estimated the absolute 
visual magnitude by fitting a theoretical zero-age horizontal branch (ZAHB) to 
the lower bound of the HB star distributions.
The absolute age of the globulars is estimated by the isochrone which, at fixed chemical
composition, best matches the CMD from the main sequence turn off (MSTO) point
to the beginning of the sub giant branch (SGB). 
Data plotted in Fig.~\ref{fig:HBRAGE} show that the HB morphology, 
at fixed cluster age, can move from very blue to very red. Indeed, 
the $HBR'$ index, within the uncertainty on the cluster age 
($\sim$ 1 Gyr, displayed by the error bar in the left bottom corner), 
ranges from 1 to 3. This evidence shows that the $HBR'$ index does 
not show any relevant correlation either with cluster age or with iron content.  

In the figure a significant fraction of GGCs is located around the two extreme 
values attained by the $HBR'$ index, namely 1 (eight globulars) and 3 (ten globulars). 
This means that the weak sensitivity of the $HBR'$ index does not allow us 
to properly trace the variation of the HB morphology when moving 
from metal-poor to metal-rich globulars, while the $HBR'$ index manages to trace the variation of the 
HB morphology in the metal-intermediate regime, where the relation appears to be almost linear. From the $HBR'$-age plane we can merely identify three 
different groups of clusters, with of $HBR'$ $\sim$ +1, $\sim$ +2, and $\sim$ +3, respectively.

%_______________________________________________________________________________

\section{L1 and L2 HB morphology indices}\label{sec:4}
A new approach to parametrize the HB properties has been recently proposed by 
\citet{milone}. They introduced two indices, L1 and L2, and provided several possible 
correlations between the HB morphology and GGC global properties.  
The L1 index measures the distance in colour between the RGB at the same 
HB magnitude level and red HB stars; the L2 index is the colour extension of the HB. L1 and L2 are very easy to estimate, since their definition relies on the selection 
of three different points on an optical ($m_ {F606W}$, $m_ {F606W}$-$m_ {F814W}$) CMD. Since they are based on colour differences, they are independent of both cluster distance and 
reddening. Moreover, empirical evidence indicates that L1 and L2 correlate with 
intrinsic GGC properties. 

Despite these advantages, the identification of the reddest HB point might be affected by the 
contamination of field stars since they attain similar colours. Moreover, the GGCs for 
which the reddest HB stars are RR Lyrae stars require additional information concerning 
their mean magnitudes. Furthermore, there is mounting evidence that the HB morphology
changes when moving from the innermost to the outermost cluster regions 
\citep{castellani,iannicola,milone}. We also note that the perceived
values of both L1 and L2 might be also affected by differential reddening.
%%%%Figures%%%%%%%%%%%%
%FIGURE 7
\begin{figure}
\centering
\includegraphics[width=\hsize]{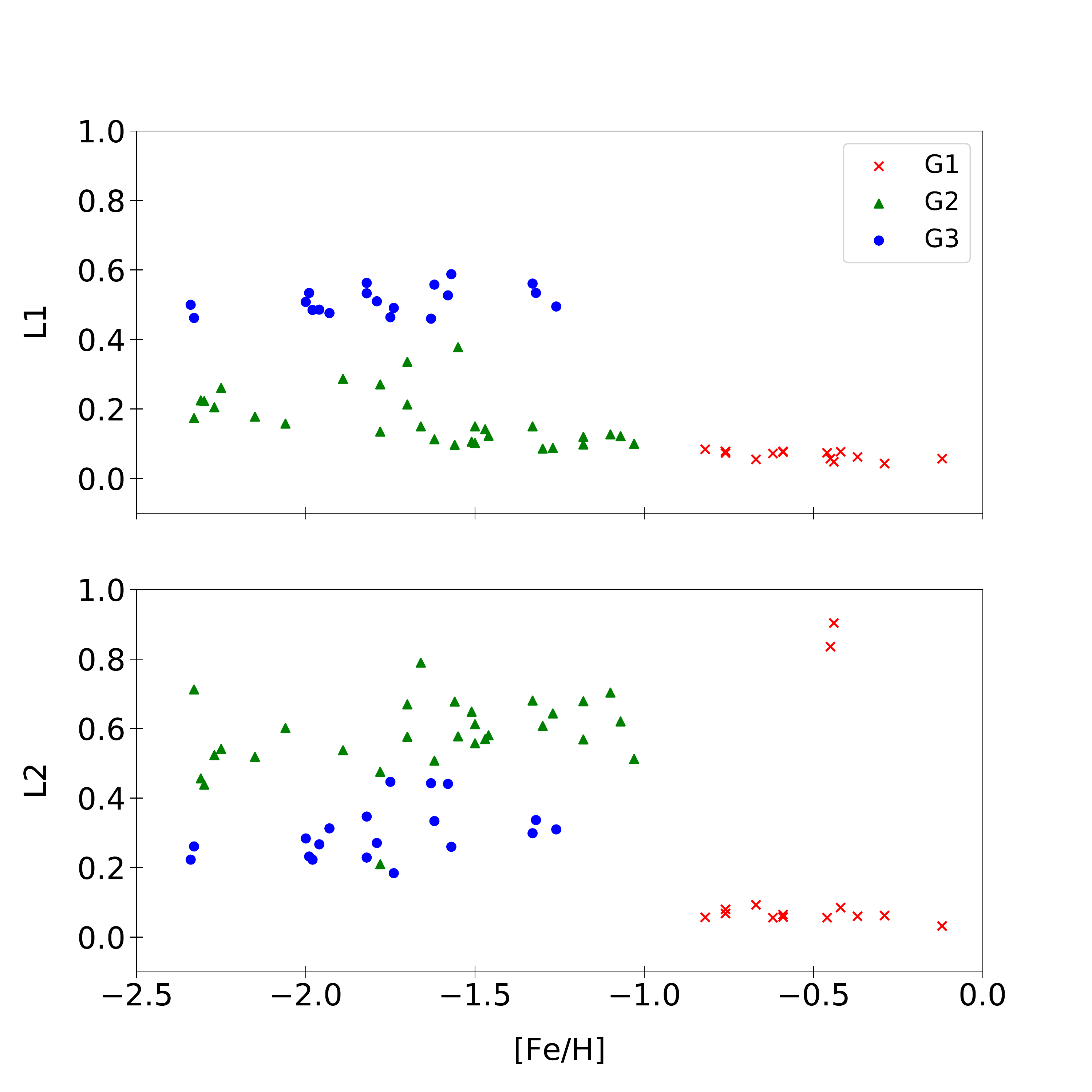} 
\caption{L1 and L2 indices as a function of the metal content \citep{carretta}. 
The different symbols and colours identify the different 
cluster groups defined in \citet{milone}: G1 (red crosses) for metal-rich
globulars ([Fe/H]>-1.0), G2 (green triangles) for clusters with [Fe/H]<-1.0
and L1 $\leq$ 0.4, G3 (blue circles) for globulars with L1 $\geq$ 0.4.}
\label{fig:Lfeh}
\end{figure}

Figure~\ref{fig:Lfeh} \citep[similar to Fig.~2 in][]{milone} shows the new 
indices as a function of GGC metallicity \citep[from][]{carretta} 
for the 62 GGCs in common with our sample. The GGC data set adopted by \citet{milone} is 
similar to the current one and relies on the same ACS@$\hst$ photometric catalogues 
we are using for the central cluster regions. This is the reason why we decided to 
use the L1 and L2 values given in Table~1 of \citet{milone} and listed in 
columns 7 and 8 of Table~\ref{table:hbr} together with their errors. 

The data plotted in the top panel of this figure show that GGCs belonging to the 
G1 and the G2 groups display a steady increase of the L1 index when moving from 
metal-rich to metal-poor clusters. The G3 clusters attain an almost
constant L1 value over more than 1~dex in metallicity. Moreover, a glance at the data shows 
that the standard deviation is modest, but the correlation does not seem to be 
mono-parametric. Indeed, there is evidence of metal-intermediate GGCs, mainly 
the G2 clusters, with the same metal content but with the L1 index changing by almost 
a factor of two.
The data plotted in the bottom panel show that the L2 index does not show 
a correlation with the metal content (first parameter) and indeed 
metal-rich clusters (G1) display a modest extent in colour, while metal-intermediate 
and metal-poor (G2+G3) clusters cover a broad colour range. 
There are only two exceptions:  \object{NGC 6388} and \object{NGC 6441}, the two peculiar metal-rich 
globulars that host red HB and blue, extreme HB stars, plus RR Lyrae stars with unusually long periods \citep{pritzl03}. 
However, the data 
plotted in the bottom panel do not display any relevant correlation with the metal 
content. 
 
%FIGURE 8
\begin{figure}
\centering
\includegraphics[width=\hsize]{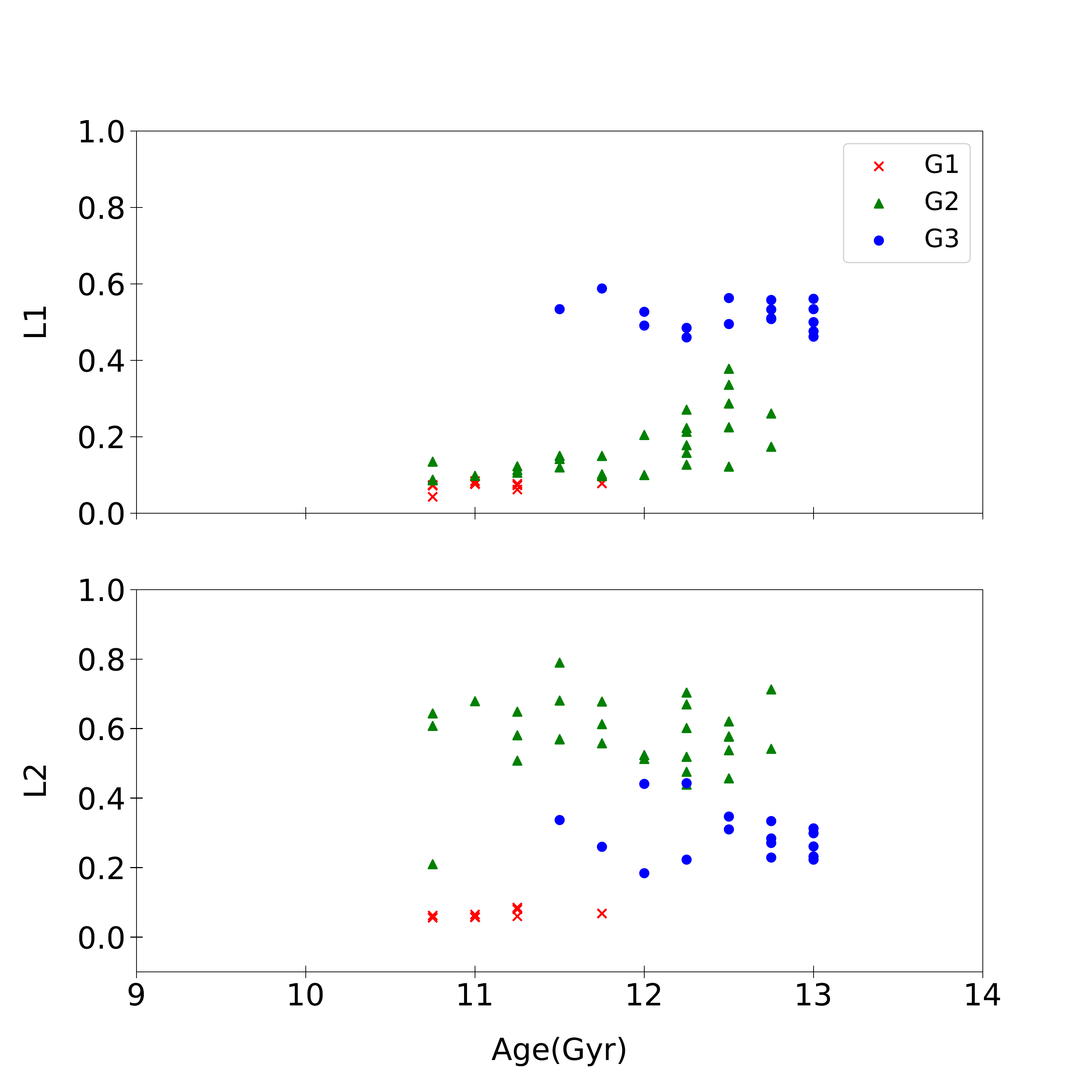} 
\caption{L1 and L2 indices as a function of the cluster 
age (Gyr) from \citet{vand,leaman}. 
The different symbols and colours identify the different 
cluster groups defined in \citet{milone}: G1 (red crosses) for metal-rich
globulars ([Fe/H]>-1.0), G2 (green triangles) for clusters with [Fe/H]<-1.0
and L1 $\leq$ 0.4, G3 (blue circles) for globulars with L1 $\geq$ 0.4.}
\label{fig:Lage}
\end{figure}

Figure~\ref{fig:Lage} \citep[similar to Fig.~5 in][]{milone} shows the correlation between 
the new L1 (top panel) and L2 (bottom panel) indices and the homogeneous cluster age 
estimates provided by \citet{vand,leaman}. The data plotted in the top panel 
seem to suggest a 
possible mild correlation between the G2 clusters and the absolute age. On the other hand, 
the G1 and the G3 clusters do not show any significant correlation with cluster age. 
The data plotted in this diagram clearly show that GGCs with the same age have L1 indices that 
change by a factor of two or three. 
The data plotted in the bottom panel display that the L2 index is independent of 
cluster age. L2 index hardly changes inside the range in age covered by G1, G2 and G3 globulars.
We performed the same analysis using the GCC ages provided by \citet{salarisweiss} and 
found similar trends.
The reader interested in a more detailed analysis of the correlation between the L1/L2 indices 
and the GGC parameters is referred to \citet{milone}.

%_____________________________________________________________
\section{A new HB morphology index: $\tau_{HB}$}\label{sec:5}

To provide a new perspective on the investigation of the HB morphology and to overcome 
some of the limitations affecting the current HB morphology indices, we have 
devised a new parameter, christened $\tau_{HB}$, based on the ratio of the areas
subtended by the CNDs as a function of magnitude ($I$-band) and colour ($\vmi$) of HB stars. To estimate the new HB morphology index first we defined a box in the $I$, $\vmi$ 
CMD large enough to include all cluster HB stars, independently of the HB morphology. Then for each cluster we computed the CND in both magnitude and in colour. 
The reference luminosity level we selected from the Harris catalogue is the 
$V$(HB), which is the mean visual magnitude of HB stars inside the RR Lyrae instability strip. 
This $V$ magnitude was transformed into an $I$-band magnitude assuming a 
mean $\vmi$ colour for the RR Lyrae stars equal to 0.45 \citep{braga}. 
To perform homogeneous 
star counts ranging from EHB to RHB stars, we defined a box covering 1.7 mag in $\vmi$ colour and
6.5 mag in $I$-band, which is 4.5 magnitudes fainter and 2.0 magnitudes brighter than the $I$-band HB luminosity 
level. Finally, the selected HB stars are sorted as a function of the $I$-band apparent magnitude 
and we cumulated the star counts starting from the faintest and reddest stars in the box.

\begin{figure}
   \centering
   \includegraphics[width=\hsize]{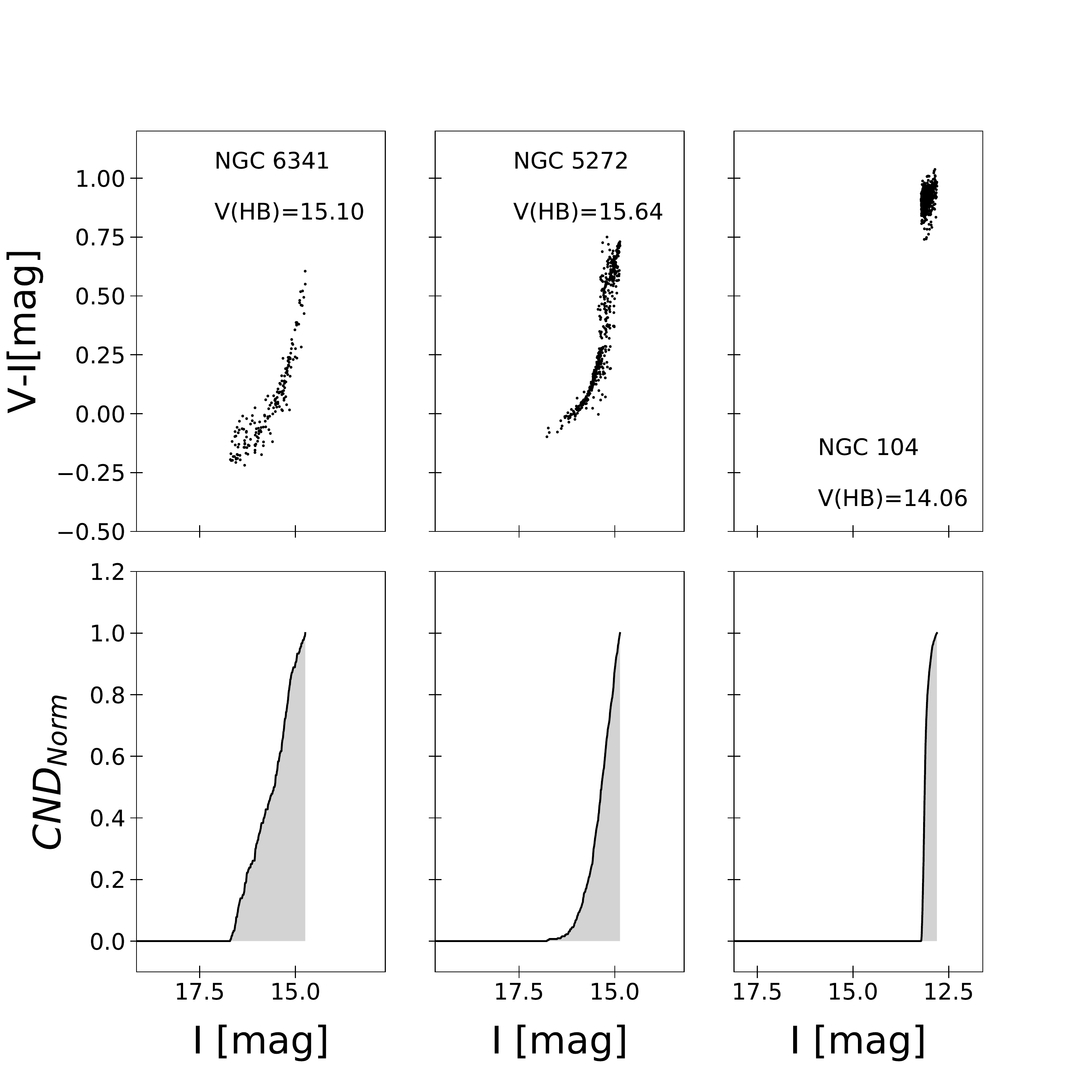}
      \caption{Run of the normalized CND with respect to the $I$ magnitude for three globular clusters (lower panels), 
chosen as representative of three different metallicity regimes (NGC~6341, NGC~5272, and NGC~104). 
Upper panels show the horizontal branch of the three chosen globulars in $I$,$\vmi$ CMDs.}
         \label{conteggiI}
   \end{figure}

The top panels of Fig.~\ref{conteggiI} display HB stars in the $I$,$\vmi$ CMD for three 
GGCs 
covering a broad range of iron abundances and HB morphologies. They are \object{NGC 6341} 
([Fe/H]=-2.25), with an HB dominated by hot and extreme blue HB stars, 
\object{NGC 5272} ([Fe/H]=-1.50) with a HB that includes blue and red HB stars,  
as well as RR Lyrae stars, and finally \object{NGC 104} ([Fe/H]=-0.76) which is 
characterized by a red HB morphology. The bottom panels of the same figure 
display the normalized $I$-band CNDs of the HB stars plotted in the top panel. 
The data plotted in these panels show that the area subtended by the three 
CNDs ($A_{CND}$, shaded area) changes significantly when moving from an HB morphology 
dominated by hot and extreme HB stars, to a morphology dominated by red 
HB stars. We find a slightly constant decrease of the $I$-band $A_{CND}$ by a 
factor of $\sim$ 1.4  when moving from NGC~6341 ($A_{CND}(I)$= 424.2) to 
NGC~5272 ($A_{CND}(I)$=336.6) to NGC~104 ($A_{CND}(I)$=217.9). A decrease on 
the order of approximately two when moving from an extreme blue HB morphology
(NGC~6341) to a red one (NGC~104).

\begin{figure}
   \centering
   \includegraphics[width=\hsize]{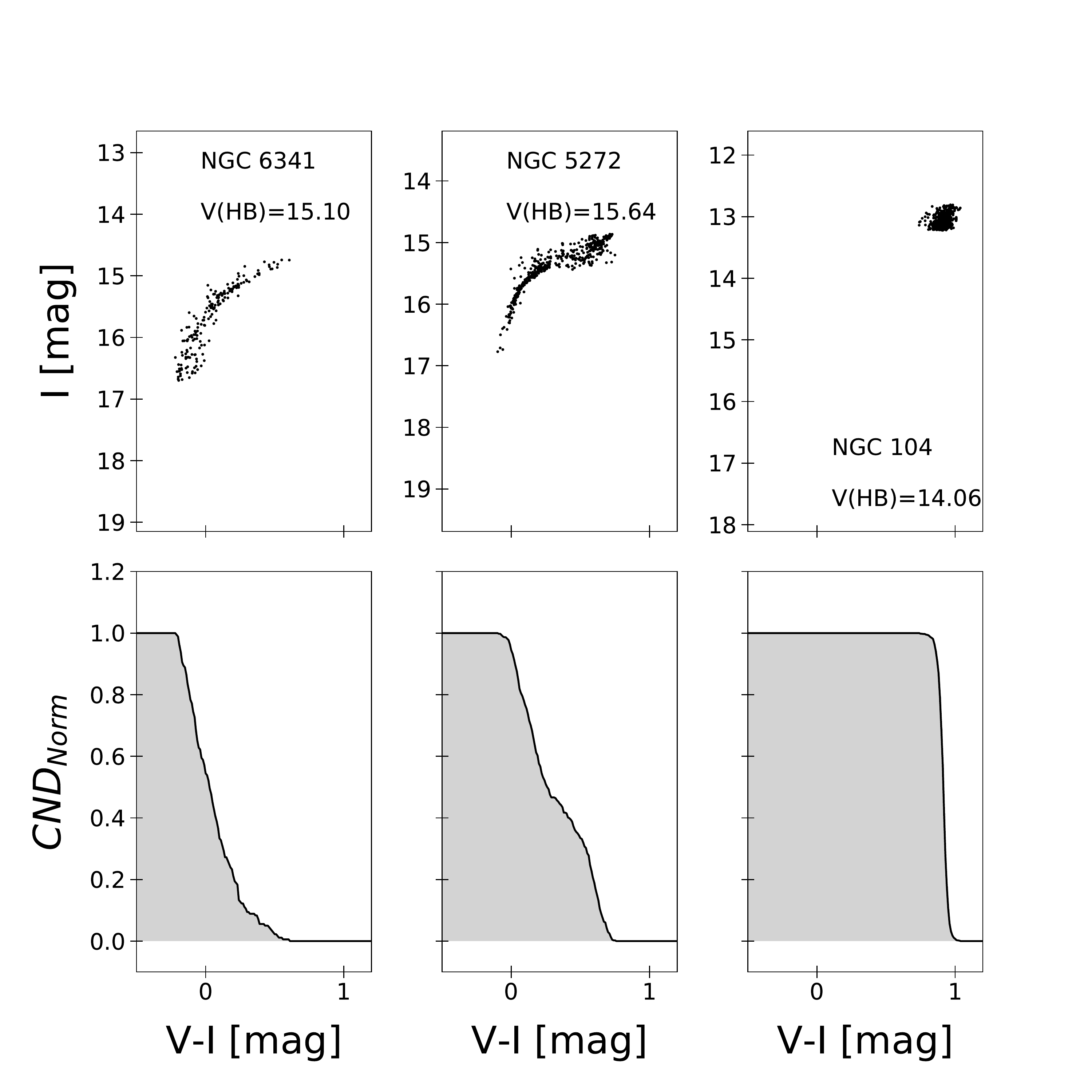}
      \caption{Run of the normalized CND with respect to $\vmi$ for the same three globular clusters 
(lower panels) of Fig. \ref{conteggiI}. 
Upper panels show the horizontal branch of the three chosen globulars in $I$, $\vmi$ CMDs.}
         \label{conteggiVI}
   \end{figure}

Then, we sorted the HB stars located inside the box as a function of the 
($\vmi$) colour. The apparent colours were de-reddened using the 
colour excesses (E($B-V$)) listed in Table~\ref{table:gcsuni}. In this case the 
HB stars were counted starting from the reddest (coldest) HB stars. 
The top panels of Fig.~\ref{conteggiVI} display HB stars in the $I$, $\vmi$ CMD for 
the same GGCs as in Fig.~\ref{conteggiI}. The bottom panels display the normalized 
$\vmi$ CND of the HB stars plotted in the top panel. The shaded area shows 
a stronger sensitivity compared with the $I$-band $A_{CND}$; indeed, they increase 
by almost a factor of three when moving from NGC~6341 ($A_{CND}(\vmi$)= 47.4) to 
NGC~5272 ($A_{CND}(\vmi$)=81.5) and to NGC~104 ($A_{CND}(\vmi$)=137.9).  

To estimate the CNDs in magnitude and in colour we accounted for the 
entire sample of HB stars, meaning stars located inside the RRL instability strip 
were included. The spread in magnitude and in colour of these objects is larger 
compared with typical HB stars. The difference is intrinsic, since we are 
using the mean of the measurements in both the $V$ and the $I$ band. In other words,
we did not perform an analytical fit of the phased  measurements.

\begin{figure}
   \centering
   \includegraphics[width=\hsize]{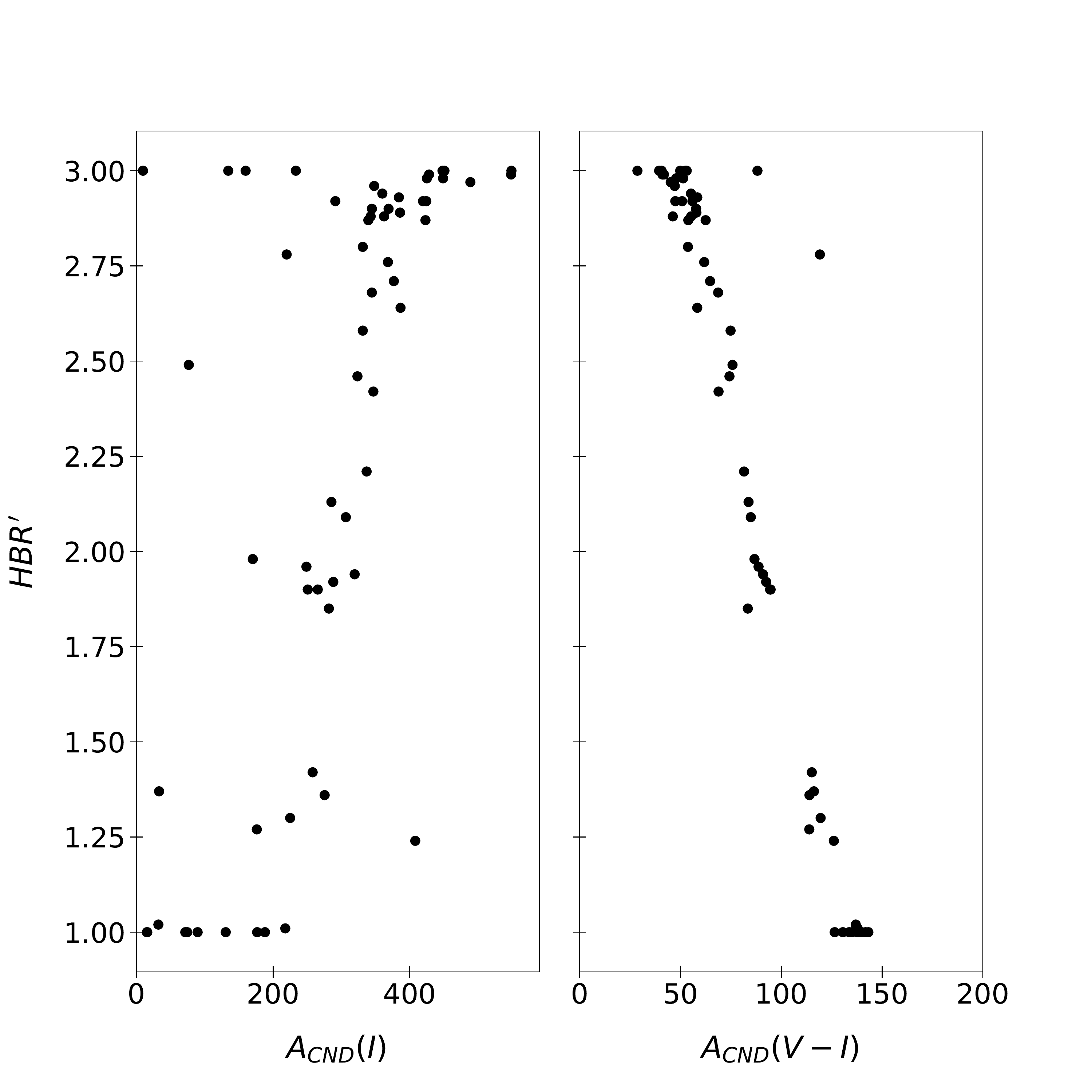}
   \caption{The indices $A_{CND}(I)$ and $A_{CND}(\vmi)$ vs. the $HBR'$ morphology index.}
   \label{fig:INTSHBRp}
 \end{figure}

To investigate the difference between the classical
($HBR'$) and the new ($\tau_{HB}$) morphology index on a more quantitative basis, the left panel of 
Fig.~\ref{fig:INTSHBRp} shows the comparison between the $I$-band $A_{CND}$ and 
the $HBR'$ index. The data plotted in this panel display a very well-defined 
linear correlation between $I$-band $A_{CND}$ and $HBR'$ index in the regime in which 
the latter attains intermediate values. The saturation for clusters dominated 
by blue HB morphologies ($HBR'\sim $ 3) and red HB morphologies ($HBR'\sim$1) is 
also quite clear at the top and the bottom of the panel. 
The right panel shows the same comparison as the left panel, but for the 
$\vmi$ $A_{CND}$. The data plotted in this panel show a linear anti-correlation over 
the entire range of values attained by the $HBR'$ index.  

We estimated the ratio between the area covered by the CND in magnitude 
and in colour, namely 
\begin{equation}
\centering
\tau_{HB}=\frac{A_{CND}(I)}{A_{CND}(\vmi)}
.\end{equation}

The $V$ magnitude levels of the HB, which we transformed to $I$(HB) using the mean $\vmi$ of 0.45~mag, 
$A_{CND}(\vmi$), 
$\tau_{HB,}$ and their errors for the GGCs in our 
sample are listed in columns 10, 11, and 12 of Table~\ref{table:hbr}, respectively. 
We propagated the error on $\tau_{HB}$ considering Poissonian uncertainties in the 
$A_{CND}$.  
%_______________________________________________________________________________________
% FIGURE 9
% FIGURE 10
%FIGURE 11
% FIGURE 12
\begin{figure}
\centering
\includegraphics[width=\hsize]{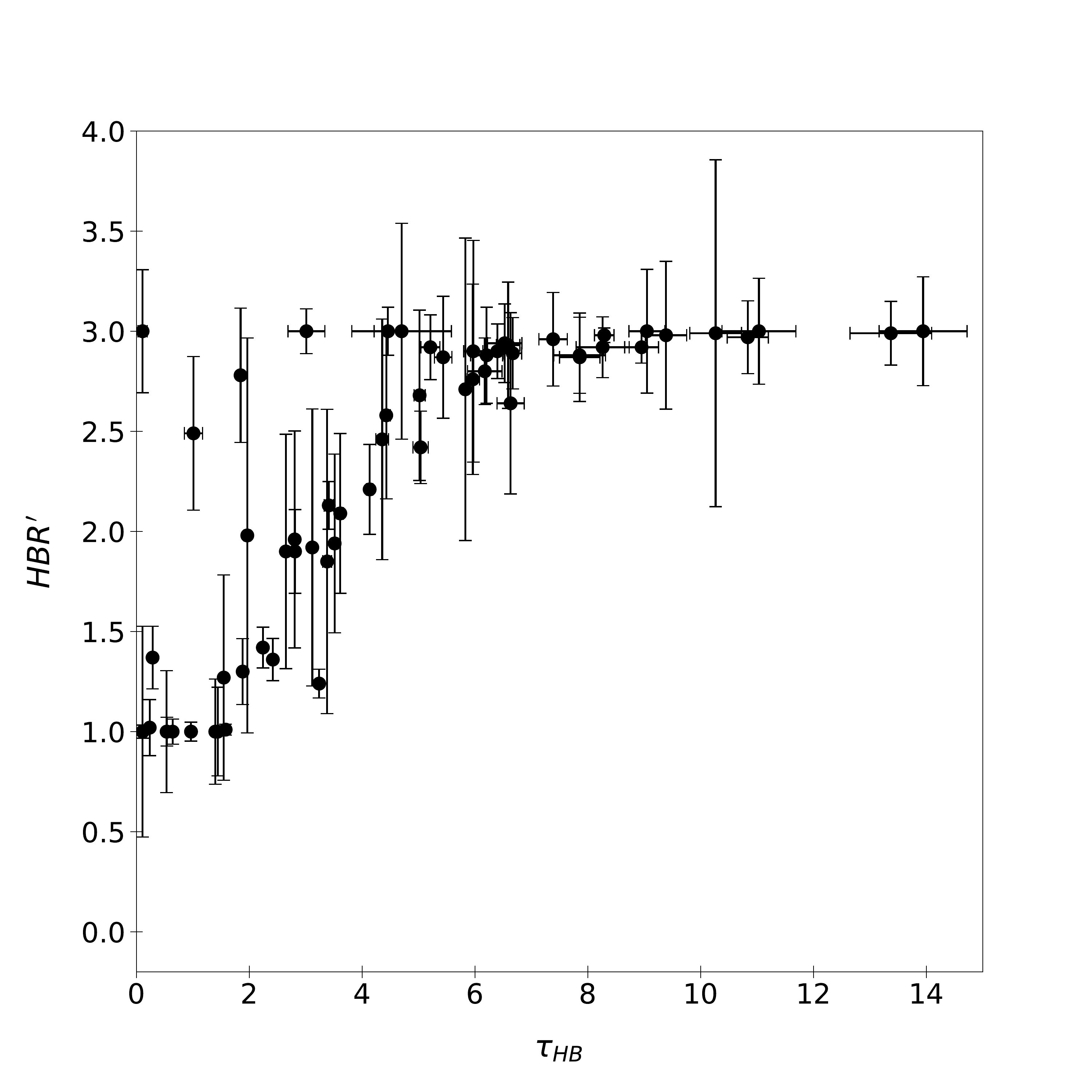}
\caption{Our new HB morphology index $\tau_{HB}$ vs. $HBR'$.}
\label{fig:TAUHBR}
\end{figure}
% FIGURE 13
\begin{figure}
\centering
\includegraphics[width=\hsize]{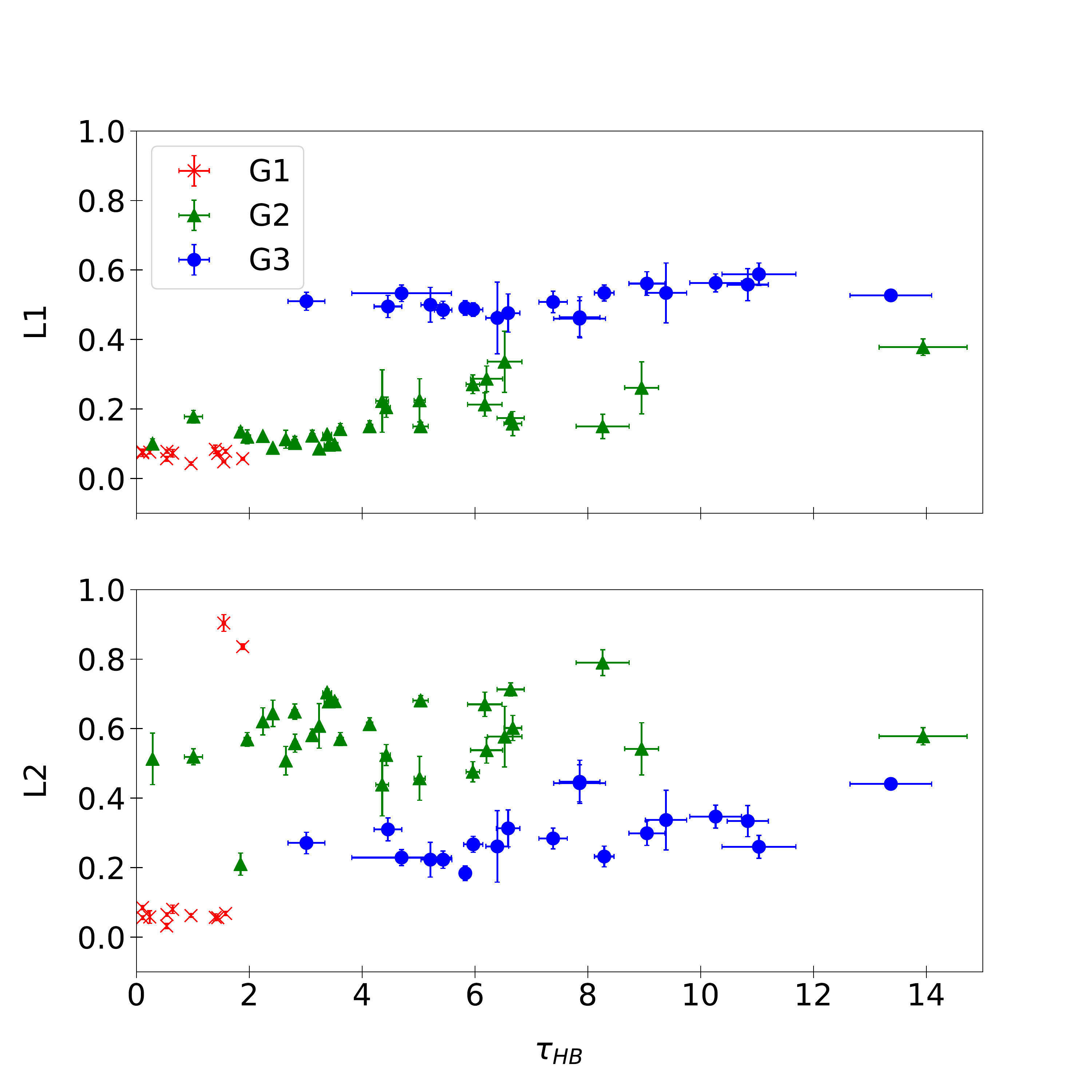}
\caption{L1 (top) and L2 (bottom) indices as a function of  
$\tau_{HB}$ for the globulars in our sample. The different symbols and colours identify the different 
cluster groups defined in \citet{milone}: G1 (red crosses) for metal-rich
globulars ([Fe/H]>-1.0), G2 (green triangles) for clusters with [Fe/H]<-1.0
and L1 $\leq$ 0.4, G3 (blue circles) for globulars with L1 $\geq$ 0.4.}
\label{fig:TAULs}
\end{figure}

%_______________________________________________________________________________
Although $\tau_{HB}$ estimate is slightly more complex than the other 
indices presented in the literature, the new HB morphology index presents several key advantages compared with the 
classical $HBR'$ and L1/L2 indices: 

{\it Dynamical range} --- The current sample covers a range in $\tau_{HB}$ that is 
a factor of seven larger than the range covered by the $HBR'$ index. The data plotted in Fig.~\ref{fig:TAUHBR} show that when 
moving from metal-rich to less metal-rich GGCs (-1.0 < [Fe/H]<0.) the 
new index changes from zero 
to roughly three while the old one changes 
from one to roughly 1.5.
The correlation between the old and new indices is linear in the metal 
intermediate regime, but it degenerates in the more metal-poor 
regime. Indeed, the $\tau_{HB}$ index increases by 
more than a factor of two (from six to fourteen), whereas the 
$HBR'$ takes on an almost constant value of three. 

The variation in $\tau_{HB}$ is a factor of twenty larger than 
the range covered by the L1 and L2 indices. The data plotted in 
Fig.~\ref{fig:TAULs} show that the G3 group is characterized by 
L1 values that are almost constant (L1 $\sim$ 0.5), while 
$\tau_{HB}$ changes from three to thirteen. The G1 and G2 groups 
display a mild linear correlation between L1 and $\tau_{HB}$, 
but once again the variation in the L1 index is modest when 
compared with $\tau_{HB}$ index (0.3 vs. 7). The correlation 
between L2 and $\tau_{HB}$ is more noisy with a large spread at 
fixed $\tau_{HB}$ value.

{\it Detailed sampling of the HB morphology }-- 
The CND in both magnitude and colour is sensitive to 
the star distribution along the HB. Data listed in Table~\ref{table:hbr} show that 
two metal-poor clusters, \object{NGC 4833 }and  NGC~6341, with similar $HBR'$ 
(2.88$\pm$0.07 vs. 2.92$\pm$0.07) and L1 (0.287 $\pm$ 0.037 vs. 0.261 $\pm$ 0.075) indices,
attain $\tau_{HB}$ values that differ at the 50\% level (6.21$\pm$0.21 vs. 8.95$\pm$0.30). 
The same applies to metal-rich clusters, and indeed, two clusters like NGC~104 and  
\object{NGC 6637} have the same $HBR'$ (1.01$\pm$0.04 vs. 1.00$\pm$0.09) and L1
(0.078 $\pm$ 0.005 vs. 0.078 $\pm$ 0.004) values, but the $\tau_{HB}$ value
of the former cluster is a factor of three larger than the latter one 
(1.58$\pm$0.01 vs. 0.54$\pm$0.02). Data listed in Table~\ref{table:pairs} highlight several other pairs of GGCs characterized by a very similar iron content (column 2), 
$HBR'$ (column 3), and L1\footnote{We did not consider L2 since 
it does not correlate with age, metallicity, and $\tau_{HB.}$} (column 4) value, 
but quite different $\tau_{HB}$ values (column 5). The various pairs are
separated in the table with horizontal lines. 
\setcounter{table}{2}
\begin{table}
\scriptsize
\caption{Examples of globular pairs of similar metallicity and $HBR'$, but different values in $\tau_{HB}$.}           
\label{table:pairs}      
\centering  
\begin{tabular}{c c c c c}  % 5 columns
\hline
\hline
 ID  & [Fe/H] & $HBR'$ & L1 & $\tau_{HB}$ \\
\hline
NGC~0104  & -0.76 $\pm$ 0.02 & 1.01 $\pm$ 0.04 & 0.078 $\pm$ 0.005 &1.58 $\pm$0.01 \\

NGC~6652  &-0.76 $\pm$ 0.14 &1.00 $\pm$ 0.28 & 0.073 $\pm$ 0.011 & 0.64 $\pm$ 0.01 \\

\hline
NGC~6121  & -1.18 $\pm$ 0.02 & 1.98 $\pm$ 0.01 & 0.120 $\pm$ 0.020 & 1.96 $\pm$ 0.06 \\

NGC~1851  & -1.18 $\pm$ 0.08 & 1.95 $\pm$ 0.01 & 0.098 $\pm$ 0.004& 3.51 $\pm$ 0.06 \\

\hline
NGC~6752 & -1.55 $\pm$ 0.01 & 3.00 $\pm$ 0.08 & 0.378 $\pm$ 0.024 & 13.94 $\pm$ 0.78 \\

NGC~6934  & -1.56 $\pm$ 0.09 & 2.13 $\pm$ 0.02 & 0.097 $\pm$ 0.0013 & 3.41 $\pm$ 0.08 \\

\hline
NGC~6144 & -1.82 $\pm$ 0.05 & 3.00 $\pm$ 0.18& 0.533 $\pm$ 0.024 & 4.98 $\pm$ 0.88\\

NGC~6541 & -1.82 $\pm$ 0.08 &2.99 $\pm$ 0.07 & 0.563 $\pm$ 0.026 & 10.26 $\pm$ 0.46\\
\hline
\end{tabular}
\end{table}

{\it Global star count of HB stars }-- The $\tau_{HB}$ does not require any identification 
of specific subgroups (blue, red, variables).   
%_______________________________________________________________________________
\section{Comparison between space- and ground-based data}\label{sec:6}
%________________________________ 
% FIGURE 14
\begin{figure}
\centering
\includegraphics[width=\hsize]{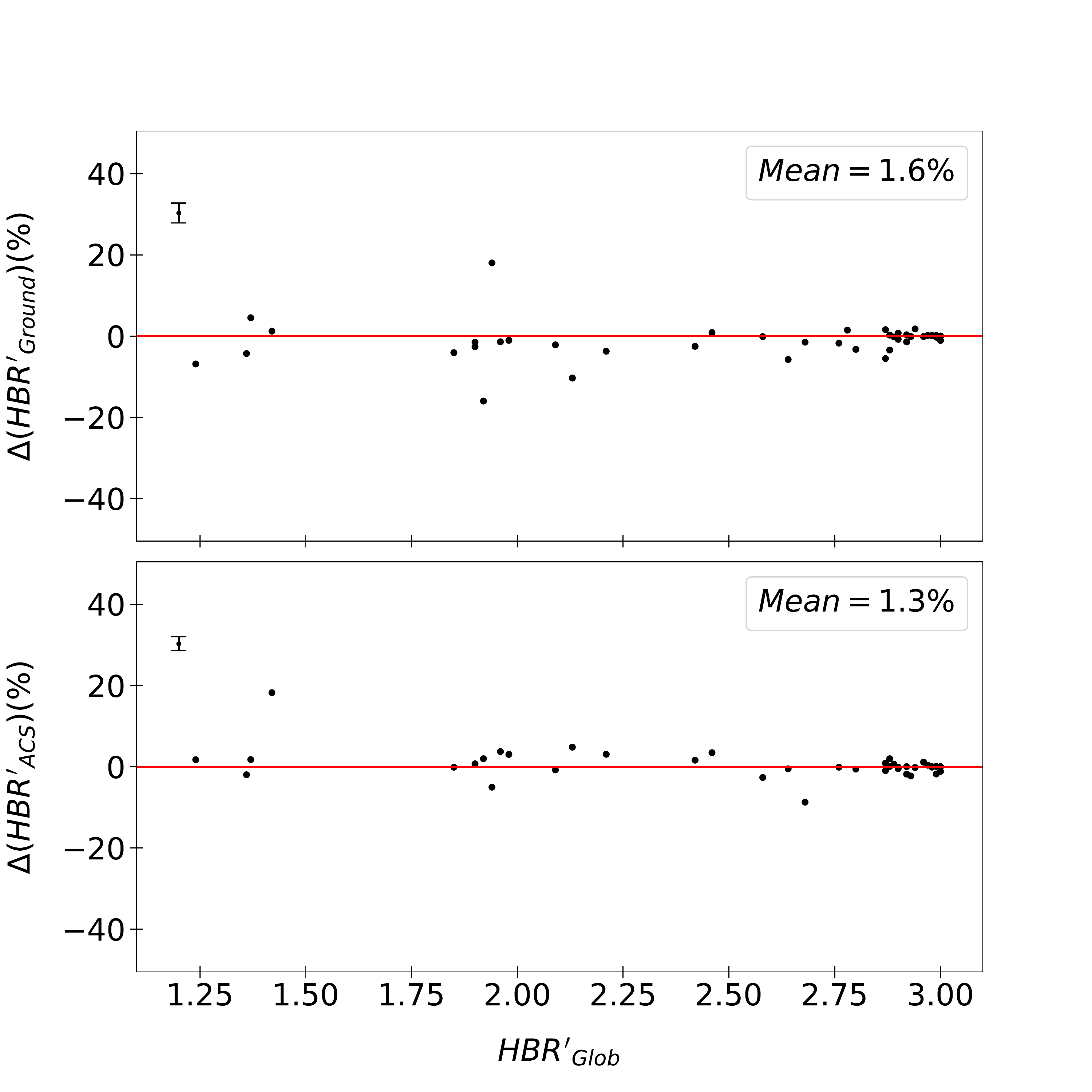}
\caption{Top: relative difference between the global $HBR'$ and 
the $HBR'$ values only based on ground-based data versus the global index. 
Note that in the estimate of the global index the priority in selecting the photometry
was given to space-based (ACS at $\hst$) data.   
Bottom: as the top panel, but the relative difference is between the global 
index and the $HBR'$ index only based on space-based data. The standard
deviation of the estimates is represented by the error bar at the top left corner
of the panels. At the top right corners we display the mean relative difference. }
\label{fig:pdiffhbr}
\end{figure}
%________________________________ 

%________________________________ 
% FIGURE 15
\begin{figure}
\centering
\includegraphics[width=\hsize]{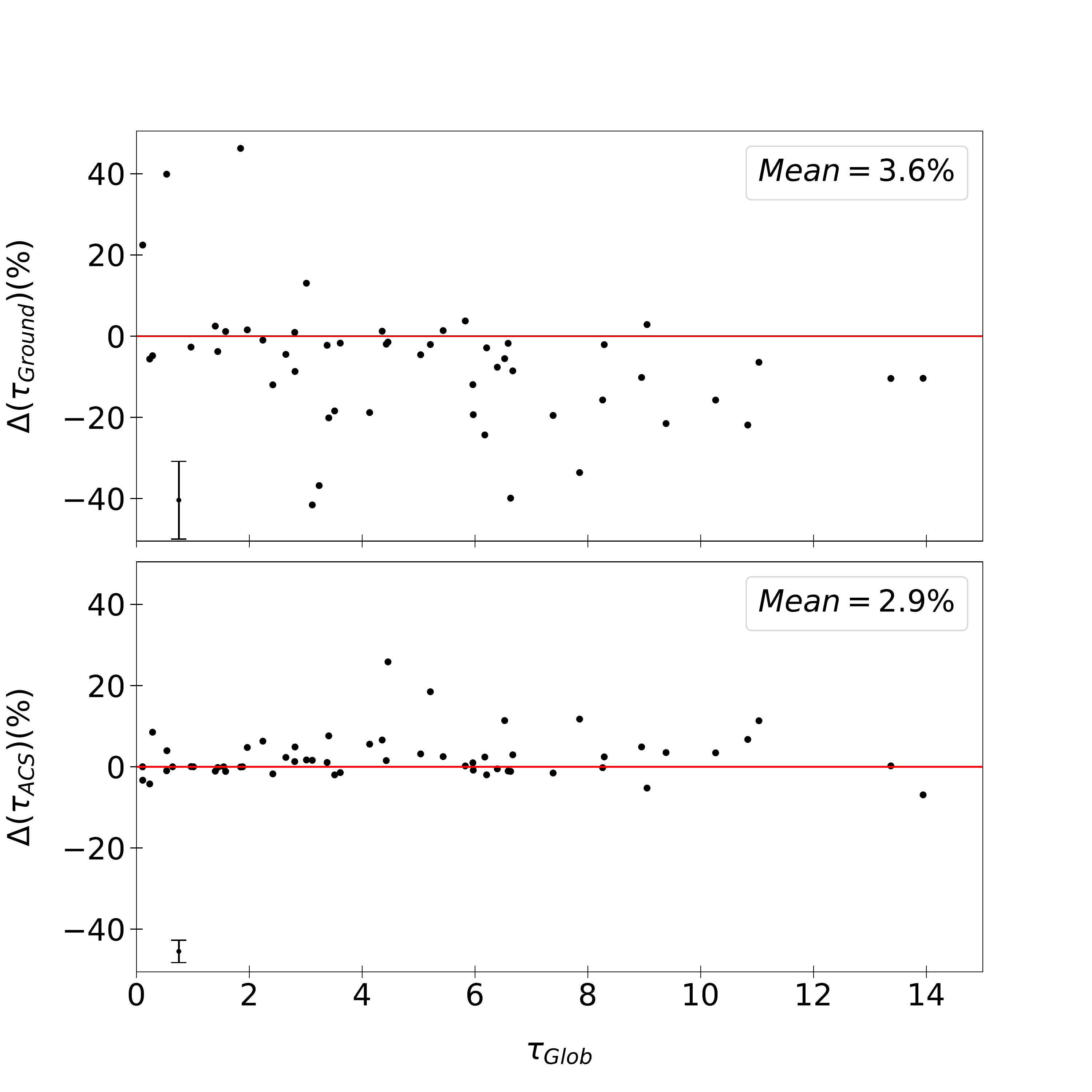}
\caption{Top: relative difference between the global $\tau_{HB}$ and the 
$\tau_{HB}$ index only based on ground-based data versus the global index.  
Note that in the estimate of the global index the priority in selecting stars was given to 
space-based (ACS at $\hst$) data.
Bottom: as the top panel, but the relative difference is between the global 
index and the  $\tau_{HB}$ index only based on space-based data. The standard
deviation of the estimates is represented by the error bar at the bottom left corner
of the panels. At the top right corners we display the mean relative difference. }
\label{fig:pdifftau}
\end{figure}
%_______________________________________________________________________________ 
As described in Sect.~\ref{sec:2}, when available we preferred the space-based  
observations for our analysis, while we used the ground-based data
for the outer regions of the globulars in our sample. 
In this section, we analyse the $HBR'$ and $\tau_{HB}$ index evaluations  
using either ACS or ground-based observations, comparing them to the `global' index, 
which is the index estimated from both space- and ground-based data
(see Sects.~\ref{sec:3} and~\ref{sec:5}), giving higher priority to the first ones when both measures where available. 

Figure~\ref{fig:pdiffhbr} shows the relative difference between 
the $HBR'$ index estimated by using either ground-based (top panel) or ACS (bottom panel) data 
as a function of the global value listed in column 6 of Table~\ref{table:hbr}.
Data plotted in this figure show the following results.
The top panel of Fig.~\ref{fig:pdiffhbr} shows that the relative difference in 
$HBR'$ index from ground-based data is on average approximately 1.6\%. 
The major exceptions are represented by the globulars NGC~1851 ($|\Delta(HBR'_{Ground})| 
\sim 18\%$) and \object{NGC 7006} ($|\Delta($HBR'$_{Ground})| \sim 16\%$). The difference 
is mainly caused by the star counts of red HB stars (R) from ground-based 
observations. They provide a lower contribution when compared with space-based 
data ($HBR'_{Ground}\sim 2.28$ vs. $HBR'_{Global}\sim 1.94$ for the former cluster, 
$HBR'_{Ground}\sim 1.61$ vs.  $HBR'_{Global}\sim 1.92$ for the latter one). 

The bottom panel shows that the difference in $HBR'$ index between $HBR'_{ACS}$ 
and $HBR'_{Global}$ is on average $\sim$1.3 \%. It is marginally lower than the 
difference based on ground-based data. 
The single exception is \object{NGC 6362}, which is characterized by $|\Delta(HBR'_{ACS})| \sim 18\%$. 
Its global $HBR'$ index turns out to be redder than the ACS one ($HBR'_{ACS}\sim 1.42$ vs. 
$HBR'_{Global}\sim 1.68$) since in the global $HBR'$ estimate the contribution of red HB stars 
is mainly given by red HB stars from ground-based observations. 

Figure~\ref{fig:pdifftau} shows the global $\tau_{HB}$ index, evaluated in Sect.~\ref{sec:5}, 
versus the relative difference between the $\tau_{HB}$ index based either on 
ground-based (top panel) or on ACS (bottom panel) observations and the global one. 
A glance at the data plotted in this figure discloses a couple of interesting findings. The top panel shows that the relative difference in $\tau_{HB}$ index based 
on ground-based data shows a larger dispersion compared to that from
space-based (bottom panel) data. The mean relative difference between 
$\tau_{HB, Global}$ and $\tau_{HB, Ground}$ is $\sim3.6\%$. 
Moreover, thirteen globulars have a difference of $|\Delta(\tau_{HB, Ground})|$ 
greater than $20\%$. In this context it is worth mentioning that amongst them 
eleven are small, concentrated globulars. Indeed, their half-mass radius 
$r_h$ \citep{harris} is entirely located inside the ACS FoV, namely 
NGC~1851 (23\%, $r_h = 0.51 \arcmin $), \object{NGC 2298} (24\%, $r_h = 0.98 \arcmin $), 
NGC~5286 (32\%, $r_h = 0.73 \arcmin $), \object{NGC 5986} (51\%, $r_h = 0.98 \arcmin $), \object{NGC 6681}(28\%, $r_h = 0.71 \arcmin $), 
\object{NGC 6779} (24\%, $r_h = 1.1 \arcmin $), \object{NGC 6934} (25\%, $r_h = 0.69 \arcmin $), NGC~7006 (71\%, $r_h = 0.44 \arcmin $),  
\object{NGC 7078} (66\%, $r_h = 1.0 \arcmin $), \object{Rup 106} (32\%, $r_h = 1.05 \arcmin $), and \object{Terzan 7} (29\%, $r_h = 0.77 \arcmin $). 
This means that the main contribution in the $\tau$ index comes from space-based data, 
while ground-based data mainly contribute for HB stars located in the cluster outskirts.
Two out of the thirteen globulars are larger, but located at large distances (true distance modulus $DM \sim 15$ mag), 
namely \object{NGC 288} (27\%, $r_h = 2.23 \arcmin $) and NGC~5272 
(27\%,  $r_h = 2.31 \arcmin $). This means that, even in these cases, 
most of the cluster stars are located inside the ACS FoV, giving a major contribution to the 
$\tau_{HB, Global}$ when compared to the ground-based observations.

The bottom panel shows that the relative variation between $\tau_{HB, Global}$ and 
$\tau_{HB, ACS}$ is, as expected low, on average 2.9\%. The only exception is \object{NGC 6717} 
(26\%), in which ACS data do not populate the brighter region of the HB, giving a 
different estimate of $\tau_{HB, ACS}$ compared to the global one (5.61 vs. 4.46).

These findings support the fact that both ground-based and space observations of our sample 
of GGCs are statistically consistent and reliable. Moreover, our estimates of $HBR'$ and $\tau_{HB}$ are valid and well grounded: independently 
of the choice about the origin of the data they give the same results within a difference 
of at most 20\% for $\tau_{HB}$ estimates.

The evidence of a larger relative difference in the $\tau_{HB}$ index shows that 
potentially our new index could quantify the different possible contribution to the HB 
morphology given by inner and outer populations within the globulars. 
This difference between 
star counts based on ground-based and on space-based data will be addressed in a future 
paper focused on the spectral energy distribution of Galactic globulars.  

%_______________________________________________________________________________
\section{Correlation of $\tau_{HB}$ with cluster metallicity, age, and helium content}\label{sec:7}
To investigate the correlation between $\tau_{HB}$ and fundamental cluster parameters,
we study its relationship to cluster metallicity, absolute age, and internal 
spread in Helium abundance.
%_______________________________________________________________________________
% FIGURE 16
\begin{figure}
\centering
\includegraphics[width=\hsize]{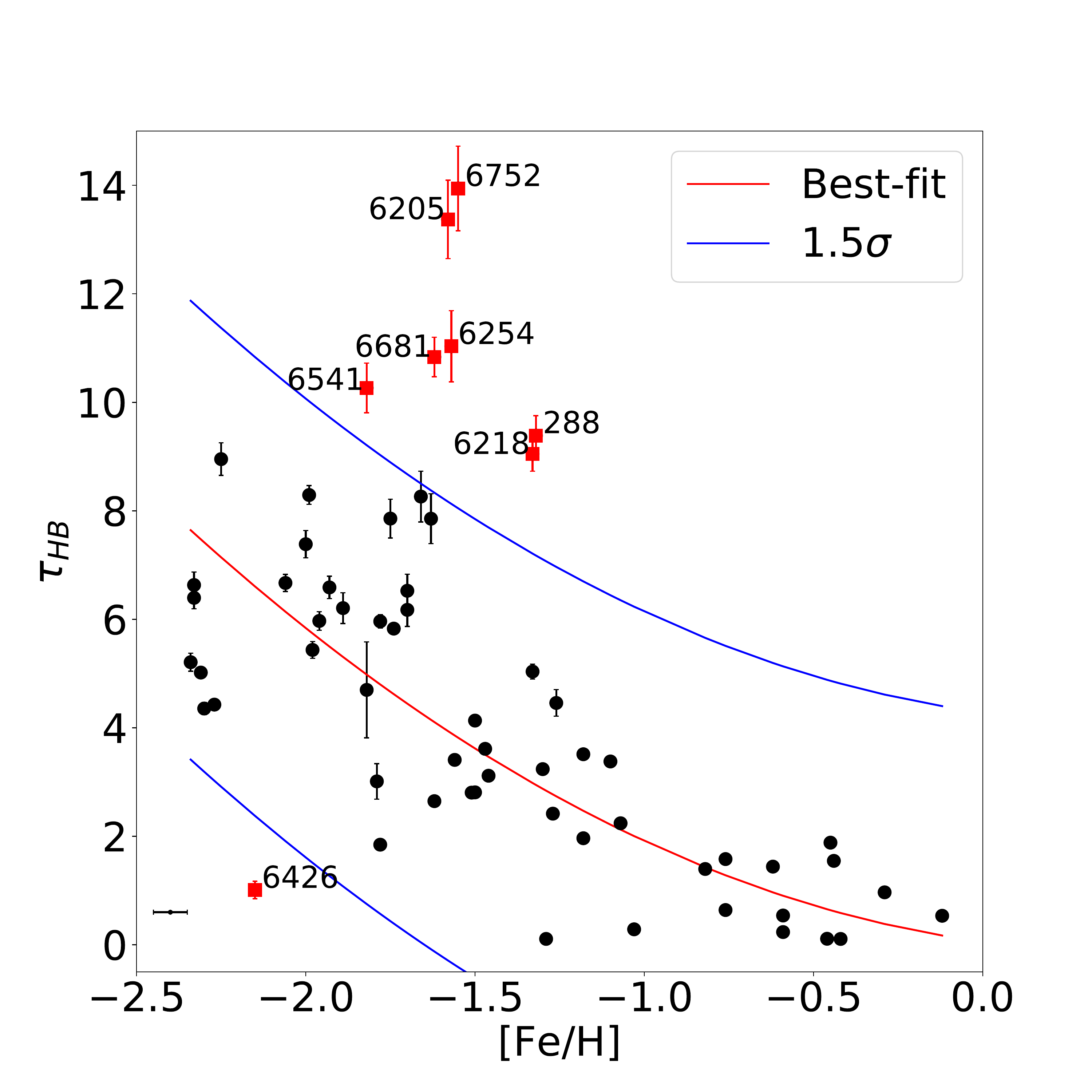}
\caption{Variation of $\tau_{HB}$ as a function of cluster metallicity \citep{carretta}. 
The red line shows the quadratic best fit, the
blue lines show 1.5 $\sigma$ levels. Red squares display the outliers, which are those
objects located at more than 1.5 $\sigma$ from the quadratic fit. 
In the left corner the error bar shows the 0.1~dex error on the metal content.}
\label{fig:TAUFEHfit}
\end{figure}
%_______________________________________________________________________________
Figure~\ref{fig:TAUFEHfit} shows $\tau_{HB}$ as a function of the iron content. 
The bulk of metal-poor GGCs ([Fe/H] $\le$ -1.5) attains 
$\tau_{HB}$ values ranging from $\sim$ 9 to $\sim$ 4. In the metal-intermediate 
regime (-1.5 $\le$ [Fe/H] $\le$ -1.0) the bulk of the GGCs are characterized by 
$\tau_{HB}$ ranging from $\sim$ 5 to $\sim$ 2, while metal-rich clusters  
([Fe/H]$\ge$ -1.0) attain $\tau_{HB}$ values smaller than 
2 on average. There is a small sample of GGCs that attains $\tau_{HB}$ values that are, 
at fixed metal content, either systematically larger or systematically smaller than the typical ranges.  
To investigate on a more quantitative basis the identification of these GGCs, 
we performed a quadratic fit of the bulk of the GGCs. We found the  best-fit function

\begin{equation}
\tau_{HB}= 0.06-0.83\cdot[Fe/H]+1.03\cdot[Fe/H]^2
,\end{equation}
 
with a $\sigma=3.08$ dispersion (see the red line in Fig.~\ref{fig:TAUFEHfit}).
We have defined as 'outliers' the GGCs that attain, at fixed metal 
content, $\tau_{HB}$ values that are more than 1.5$\sigma$ away from the
best quadratic fit.  
We ended up with a subsample of seven metal-poor and metal-intermediate
($-$1.82 $\le$ [Fe/H] $\le -$1.32) GGCs with $\tau_{HB}$ values larger 
than 9 (see red squares in Fig.~\ref{fig:TAUFEHfit}). This 
subsample includes the GGCs that attain the largest $\tau_{HB}$ values, 
namely \object{NGC 6205} ($\tau_{HB}$= 13.37 $\pm$ 0.72) and \object{NGC 6752} ($\tau_{HB}$= 13.94 $\pm$ 0.78). 
This means that they are 
characterized by very blue HB morphologies. 
Data plotted in Fig.~\ref{fig:TAUFEHfit} also show a metal-poor 
GGC (NGC~6426) with $\tau_{HB}$=1.01 $\pm$ 0.16 that is a factor of four to ten
smaller than the bulk of GGCs with similar metal abundances.   

The above empirical evidence indicates that the new HB morphology 
index appears to be a solid diagnostic to select GGCs that are 
strongly affected by the second parameter problem. To further 
investigate the nature of these clusters we decided to perform a
more detailed analysis of the GGCs in the metallicity range covered 
by the second parameter clusters.
%________________________________ CMDs SECOND PARAMETRIZED________________________
% FIGURE 17
\begin{figure*}
\centering
\includegraphics[width=18cm]{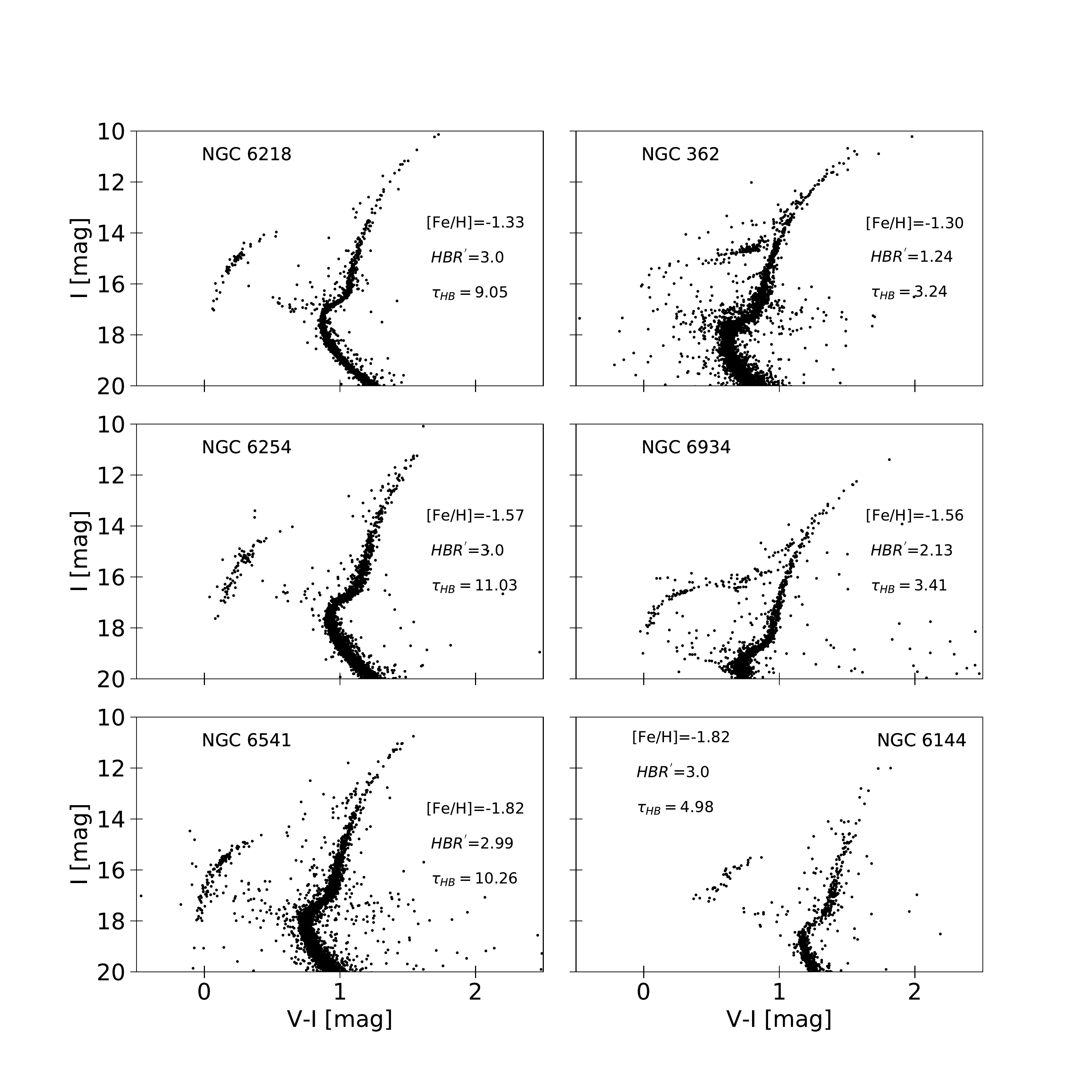}
\caption{CMDs ($I$, $\vmi$) for three pairs of GGCs in the sample. Left panels: Clusters belonging to the 
outlier group. 
Right panels: CMDs of clusters with similar [Fe/H] as the outlier ones, but following the main
$\tau_{HB}$-[Fe/H] relation.}
\label{confCMD}
\end{figure*}
%____________________________________________________________________________________
We selected three out of the eight outliers, namely \object{NGC 6218} 
([Fe/H]=-1.33, Age= 13 Gyr), \object{NGC 6254} ([Fe/H]=-1.57, Age= 11.75 Gyr),
and \object{NGC 6541} ([Fe/H]=-1.82, Age= 12.50 Gyr), covering roughly 0.5 dex
in metal content. The left panels of Fig.~\ref{confCMD} show the $I$,$\vmi$
CMDs of these clusters.
Assuming that the HB morphology is mainly driven by a difference in metal 
content, we selected three GGCs with iron abundances very similar to the selected
outliers, but with $\tau_{HB}$ values close to the best-fit line plotted in 
Fig.~\ref{fig:TAUFEHfit}, namely \object{NGC 362} ([Fe/H]=-1.30, Age= 10.75 Gyr), 
NGC~6934 ([Fe/H]=-1.56, Age= 11.75 Gyr), and \object{NGC 6144} 
([Fe/H]=-1.82, Age=12.75 Gyr). The right panels of Fig.~\ref{confCMD} display 
their $I$,$\vmi$ CMDs.
The CMDs display several interesting features worth discussing. 

The top panels of Fig.~\ref{confCMD} compare the CMDs of two   
metal-intermediate GGCs: an outlier cluster characterized 
by a very blue HB morphology, and a typical one (in terms of $\tau_{HB}$, 9.05 vs. 3.24) , 
mainly dominated by red HB stars. It is worth mentioning that the HB morphology of this pair of
clusters could also be traced on the basis of the $HBR'$ index. Indeed, the $HBR'$ value of 
NGC~6218 is a factor of 2.5 larger than the $HBR'$ value of NGC~362.

The middle panels of Fig.~\ref{confCMD} also display the CMDs of two 
metal-intermediate GGCs. The outlier GGC is characterized 
by a very blue HB morphology, while the typical one shows an HB
that hosts variable stars, blue and red HB stars. For these GGCs the 
$HBR'$ index decreases by only the 30\% when moving from NGC~6254 (3.00) 
to NGC~6934 (2.13). Interestingly enough, the $\tau_{HB}$ index differs 
by more than a factor of three.

The bottom panels of Fig.~\ref{confCMD} display the CMDs of two metal-poor GGCs. 
The outlier GGC shows, as expected, a very blue 
HB morphology, while the 'typical' one also shows a blue HB morphology.
The $HBR'$ index for these two GGCs is identical within the errors:
2.99 for NGC~6541 and 3.00 for NGC~6144. On the other hand, the 
$\tau_{HB}$ index differs by more than a factor of two (10.26 vs 4.98).  

The current findings further support the strong sensitivity of 
the $\tau_{HB}$ index to variations in HB morphology when moving from 
the metal-intermediate to the metal-poor regime. 

%____________________________________________________________________________________
%____________________ FIGURE AGE _____________________________________________________
% FIGURE 18
\begin{figure}
\centering
\includegraphics[width=\hsize]{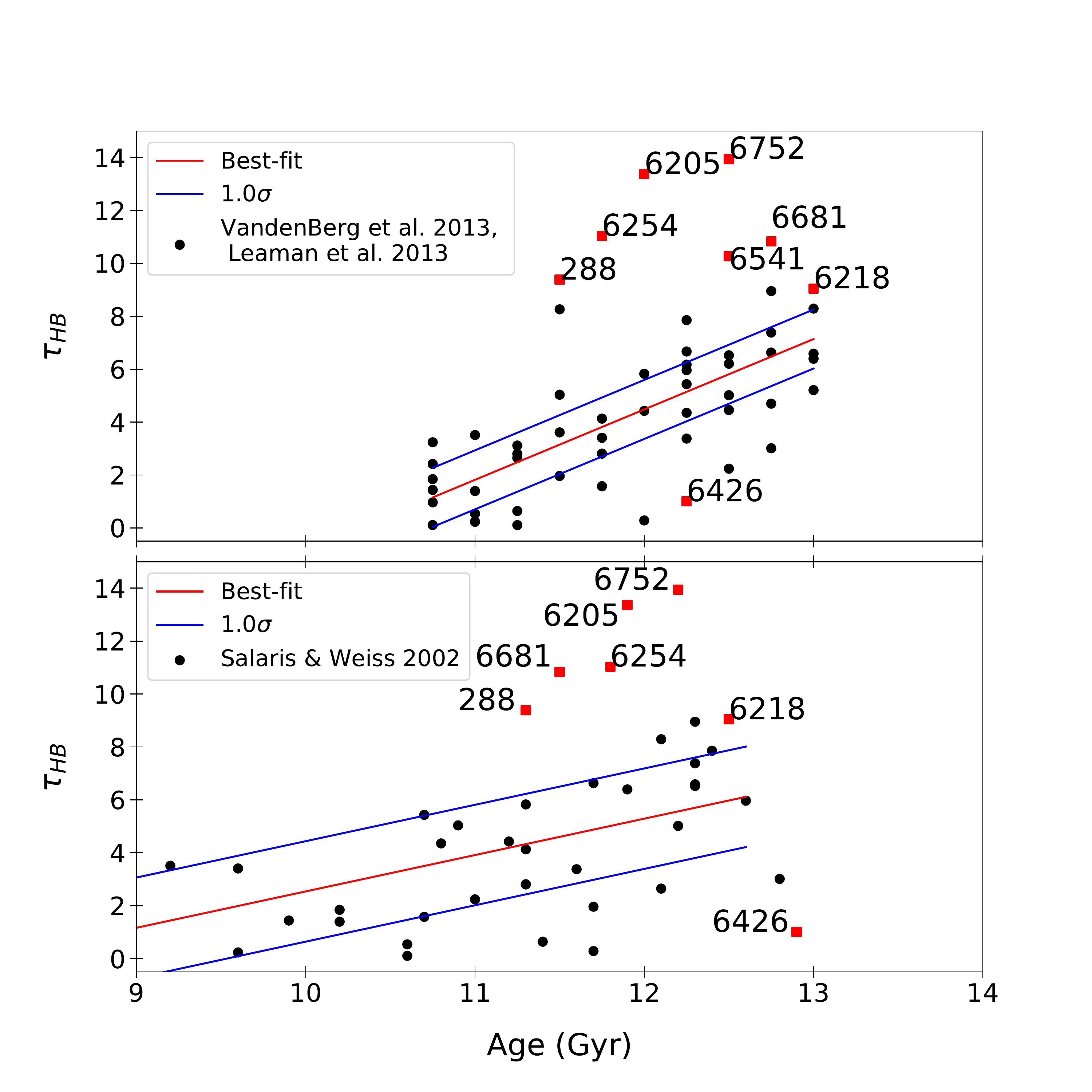}
\caption{Our index $\tau_{HB}$ as a function of cluster ages obtained from different authors. 
Top: cluster ages from \citet{vand,leaman}. The error bar in the lower left corner shows the
conservative $\pm$0.5~Gyr error on the GGC ages. Bottom: ages from \citet{salarisweiss} based on the 
metallicity scale provided by \citet{cg97}. In both panels red squares are the same second parameter clusters shown in
Fig.~\ref{fig:TAUFEHfit} and discussed in this section. Red lines determine 
the best-fit to each data sample, blue lines the $\pm 1\sigma$ levels.}
\label{fig:ages}
\end{figure}
%______________________________________________________________________________________
The anti-correlation found between the $\tau_{HB}$ index and the 
cluster metallicity does not include information about the cluster age. 
We have then investigated whether the new HB morphology index is 
correlated with the cluster ages.  

To establish the age dependence we took advantage of the recent 
homogeneous age estimates provided by \citet{vand,leaman}. The data plotted in the top panel of Fig.~\ref{fig:ages} display a well-defined linear 
correlation between age and $\tau_{HB}$. In particular, 
there is evidence that, when moving from a red to a blue HB morphology,
GGCs become on average older. We performed a linear 
fit ($\tau_{HB} =a+b \cdot Age$, red line) whose coefficients are listed in Table~\ref{table:bf},  
together with the standard deviation ($\sigma$). 
The dispersion around the linear fit is modest, equal to 1.21. However, the second parameter globulars  
identified in the $\tau_{HB}$-[Fe/H] plane do not follow the same trend. They are
on average more than 1.5$\sigma$ away from the main relation, that is,   
at fixed cluster age their $\tau_{HB}$ values are systematically larger 
than typical GGCs. This further supports the evidence that the $\tau_{HB}$ index 
is a robust diagnostic to identify second parameter GGCs. It is worth mentioning 
the presence of four GGCs (\object{NGC 6171}, NGC~6362, \object{NGC 6535}, \object{and NGC 7089}) that are 
$\sim$2.5$\sigma$ away from the linear fit, but their position might also be affected 
by uncertainties in the absolute cluster age (see the horizontal error bar plotted
in the bottom right corner).      

To further investigate the impact of possible systematics on the cluster ages, we used
the results by \citet{salarisweiss} employing the \citet{cg97} metallicity scale, for
their 43 GGCs in common 
with our sample. They divided the sample into 
four different metallicity bins and for each of them selected a calibrating 
cluster. They then estimated the absolute age of the calibrating cluster by using 
the vertical method, that is, the difference in visual magnitude ($\Delta V$) between 
the HB luminosity level and the MSTO. For all the other clusters 
in a specific metallicity bin, they estimated the relative age with respect to the 
calibrating cluster by using the horizontal method, that is, the difference 
in colour ($\Delta (\vmi)$ or $\Delta (\bmv)$) between the main sequence TO and 
the base of the RGB.

Data are shown in the bottom panel of Fig.~\ref{fig:ages}. 
Once again, we found that the $\tau_{HB}$ index correlates with the cluster age. 
We performed the same linear fit as for the age estimates by \citet{vand,leaman}, and the coefficients are listed in Table~\ref{table:bf} together with the
standard deviation, which is equal to 1.89. The red and the blue lines display the linear fit 
and the 1$\sigma$ limits, respectively. The difference between the second parameter clusters and 
the linear fit is larger than 2$\sigma$. This means that their peculiarity is independent
of the adopted absolute age. Moreover, this plot also shows a few GGCs, roughly 
2$\sigma$ away from the linear fit, already identified in the top panels: NGC~6171 and 
NGC~6535, plus a new one, NGC~6652.

\begin{table}
\scriptsize
\caption{Best fit parameters for the linear functions fitting the $\tau_{HB}$-Age relations used in this work}           
\label{table:bf}      
\centering  
\begin{tabular}{c c c c }  % 4 columns
\hline
& a &b & $\sigma$ \\
\hline
VandenBerg + Leaman et al. & -27.23 $\pm$ 6.34 & 2.69 $\pm$ 0.53 & 1.21\\
Salaris and Weiss & -11.20 $\pm$ 3.55 & 1.37 $\pm$ 0.32 & 1.89\\
\hline
\end{tabular}
\end{table}

%_______________________________________________________________________________
% FIGURE 19
\begin{figure}
\centering
\includegraphics[width=\hsize]{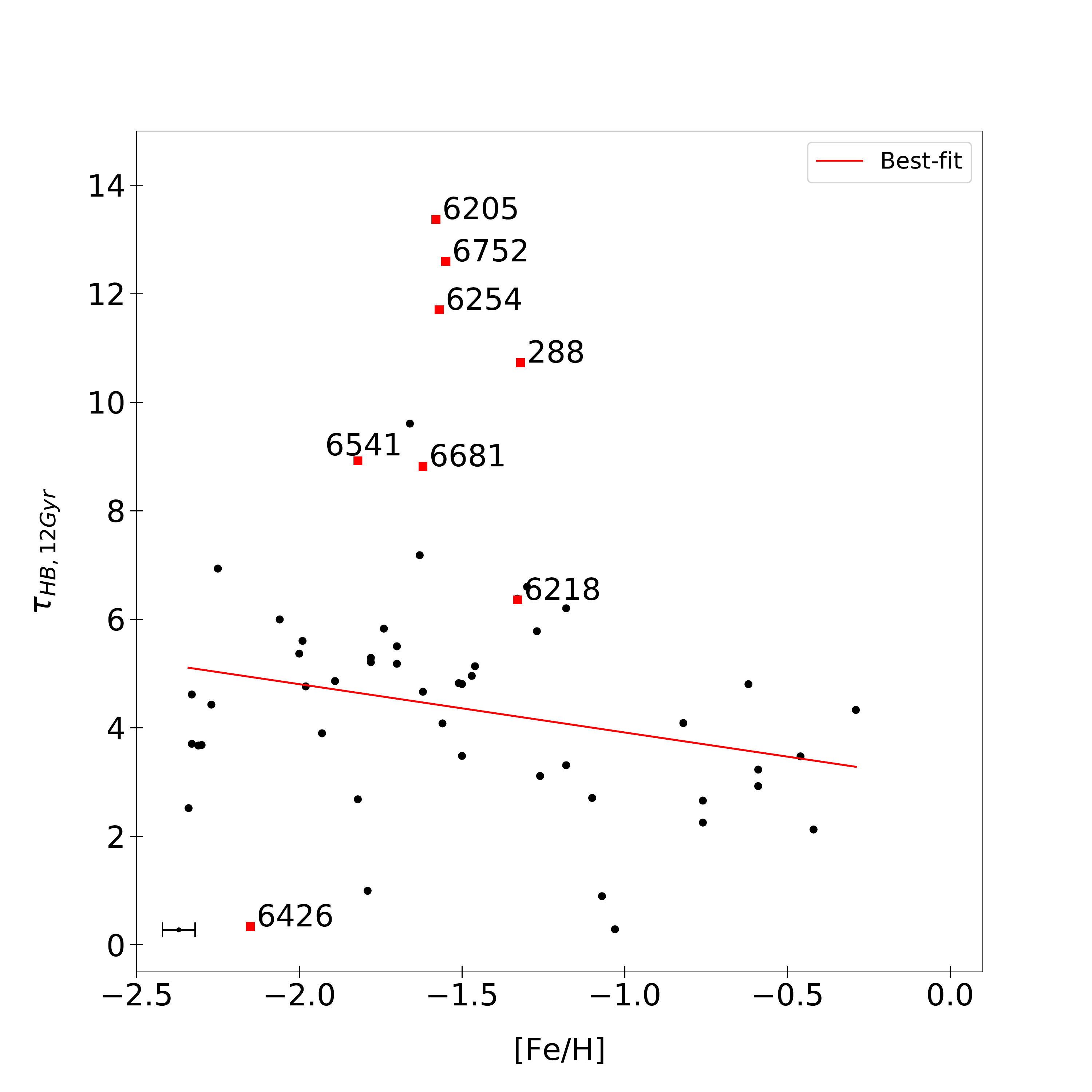}
\caption{[Fe/H] vs. $\tau_{HB, 12~Gyr}$. The red line identifies 
the linear best-fit function of the plane, and red squares mark the eight second parameter clusters (see text for details).}
\label{fig:taufehcorr}
\end{figure}
%_______________________________________________________________________________

We analysed the age and the metallicity dependencies of $\tau_{HB}$, since 
they are considered as the main culprits affecting the HB morphology. However, 
the current findings further support the need for at least one more 
parameter to explain the observed variation in HB morphology. We assumed as 
a working hypothesis that the dispersion in the $\tau_{HB}$-age diagram was 
caused by both age and helium variations \citep{dantona,caloidantona}. 
Therefore, we decided to empirically remove the age dependence of $\tau_{HB}$,
and check whether the residuals in the reduced $\tau_{HB}$-[Fe/H] 
diagram correlate with the spread in helium of the cluster. This empirical
'reduction' of $\tau_{HB}$ to a single age accounts automatically for possible
age-metallicity relations in the cluster sample, as well as for the dependence
of $\tau_{HB}$ on age at fixed metallicity. To perform this experiment we 
reduced the individual $\tau_{HB}$ measurements (Table~\ref{table:hbr}) to the
values they would have for an age of 12 Gyr \citep[$\tau_{HB, 12~Gyr}$, using
the ages by][]{vand}, as detailed below.

We first calculated the index value at an age of 12~Gyr as provided by the 
best-fit relation in the upper panel of Fig.~\ref{fig:ages} (value equal to 4.63);
then, for each cluster, we calculated the values determined from the same best-fit relation but 
for the individual cluster ages.
For each GGC we then calculated the difference $d\tau$ between the value expected 
at the cluster age and the value expected at 12~Gyr, and finally determined 
$\tau_{HB, 12~Gyr}=\tau_{HB}-d\tau$.

Figure~\ref{fig:taufehcorr} shows the $\tau_{HB, 12~Gyr}$ values as a function of [Fe/H].
In contrast to what we found in Fig.~\ref{fig:TAUFEHfit}, after eliminating on average the effect of age,
we now have a linear best-fit relation between $\tau_{HB, 12~Gyr}$ and the 
metal content (the red line in Fig.~\ref{fig:taufehcorr}):
\begin{equation}{
\tau_{HB, 12~Gyr}=3.02-0.89\cdot [Fe/H]
}
,\end{equation}
with dispersion $\sigma=1.64$. The red squares identify the second parameter clusters: they still attain $\tau_{HB, 12~Gyr}$ values far from the best-fit relation. 

%_______________________________________________________________________________
% FIGURE 20
\begin{figure}
\centering
\includegraphics[width=\hsize]{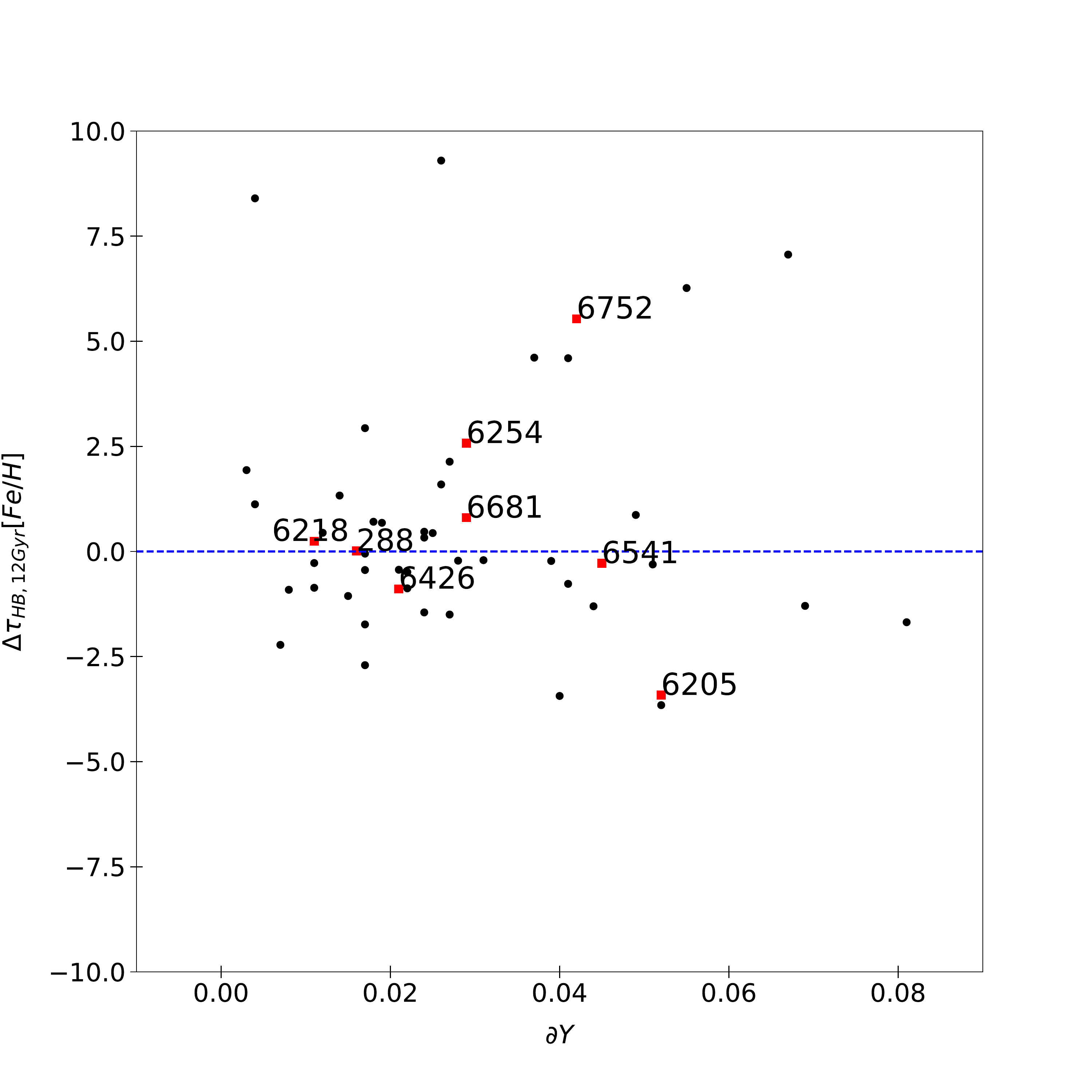}
\caption{Residuals of the corrected $\tau_{HB, 12~Gyr}$  as a function of the 
metal content \citep{carretta} versus the spread in helium content $\partial Y$ estimated 
by \citet{milone18}. Red squares identify the second parameter globulars. }
\label{fig:dtaudYcorr}
\end{figure}
%_______________________________________________________________________________

In a very recent paper, \citet{milone18} estimated the 
spread in initial helium content ($0.245\le Y\le 0.4$) for a sizeable sample of GGCs (57). They used data from the
$\hst$ UV survey of Galactic GCs \citep[$F275W$, $F336W,$ and $F438W$ 
filters of the ultraviolet and visual channel of $\hst$/WFC3 (UVIS/WFC3),][]{piotto,milone16}  and from the 
Wide Field Channel of the Advanced Camera for Surveys (WFC/ACS) 
\citep[$F606W$ and $F814W$ photometry,][]{sara,dotter11} programmes.
The He spread values, $\partial Y_{max}$, for each of the 56 GGCs in common with 
\citet{milone18} are listed in column 9 of Table~\ref{table:hbr}.

The spread in helium content ranges from almost zero for NGC~6362, NGC~6535, and NGC~6717, up to 
$\sim $0.08 for NGC~6388, NGC~6441, and NGC~7078. To further constrain the sensitivity 
of the new HB morphology index, we correlated the residuals of  $\tau_{HB,12~Gyr}$ of the best-fit function
in Fig.~\ref{fig:taufehcorr} as a function of the metallicity, shown in Fig.~\ref{fig:dtaudYcorr}. 

The figure shows that the $\tau_{HB, 12~Gyr}$ index does not seem to be correlated with the 
spread in helium content. This means that, despite our attempt to limit the age and metallicity 
effects, the spread in helium content is not able to justify the observed 
spread in $\tau_{HB}$.
We note that we also estimated  the residuals of the $\tau_{HB,12~Gyr}$ using just space data 
(see Sect.~\ref{sec:6}) and related them to $\partial Y$ values, since the latter quantities 
are evaluated from the same $\hst$ data. Once again, despite the data homogeneity, we did not find any clear correlation 
with $\partial Y$ values.  
%_______________________________________________________________________________
\section{Comparison with synthetic horizontal branch models}\label{sec:8}

To constrain on a more quantitative basis the impact that both cluster age and 
spread in helium content have on the observed spread of the new HB morphology 
index, we decided to use a novel set of synthetic horizontal branch (SHB) models.  
The synthetic HB models have been computed employing HB tracks and progenitor isochrones 
from the $\alpha$-enhanced a Bag of Stellar Tracks and Isochrones (BaSTI) stellar model library \citep[][]{bastia}
\footnote{http://www.oa-teramo.inaf.it/BASTI} and a code fully described in \citet{dale}. 
We considered three different metallicities, namely [Fe/H]=$-$0.7, [Fe/H]=$-$1.62, 
and [Fe/H]=$-$2.14 (all with [$\alpha$/Fe]=0.4). The initial He abundances of the HB
progenitors at these three metallicities are equal to $Y$=0.245, 0.246, and 0.256, 
respectively.

For each [Fe/H] we calculated first a set of 
synthetic HBs for an age equal to 12~Gyr 
(keeping $Y$ constant for each [Fe/H]), assuming the 
RGB progenitor loses an amount of mass $\Delta M=0.28 \msun$, $\Delta M=0.21 \msun,$ and
$\Delta M=0.16 \msun$ for Fe/H]=$-$0.7, [Fe/H]=$-$1.62, 
and [Fe/H]=$-$2.14, respectively. The mass loss is estimated 
with a 1$\sigma$ Gaussian spread equal to 0.01$\msun$, 
irrespective of the chemical composition. 
In addition, assuming the same age and RGB mass loss, we calculated SHBs for each metallicity 
with an uniform distribution of initial $Y$ for the progenitors, with a range $\partial Y$=0.03. 

In brief, the synthetic HB code first draws randomly a value of Y with a uniform probability distribution 
between $Y$ and $Y$+$\partial Y$ ($\partial Y$=0 for the models at constant He) and determines 
the initial mass of the star at the RGB tip ($M_{\rm TRGB}$) from interpolation amongst the 
BaSTI isochrones of the chosen age. The mass of the corresponding object evolving along the 
HB ($M_{\rm HB}$) is then calculated as $M_{\rm HB}=M_{\rm TRGB} - \Delta M$, where $\Delta M$ 
is drawn randomly according to a Gaussian distribution with the specified mean values and $\sigma$.
The magnitudes of the synthetic star are then determined according to its position along the HB track
with appropriate mass and $Y$  obtained by interpolation among the available set of HB tracks, 
after an evolutionary time $t$ has been randomly extracted. 
The value of $t$ is determined assuming that stars reach the ZAHB at a constant rate, 
employing a flat probability distribution ranging from zero to $t_{\rm HB}$, 
where $t_{\rm HB}$ is the time spent from the ZAHB to the He-burning 
shell ignition along the early asymptotic giant branch. 
The value of  $t_{\rm HB}$ is set by the HB mass with the longest lifetime 
(the lowest masses for a given chemical composition). This implies that for some 
objects the randomly selected value of $t$ will be longer than its 
$t_{\rm HB}$, meaning that they have already evolved to the next evolutionary 
stages.  

%________________________________ 
% FIGURE 21
\begin{figure}
\centering
\includegraphics[width=\hsize]{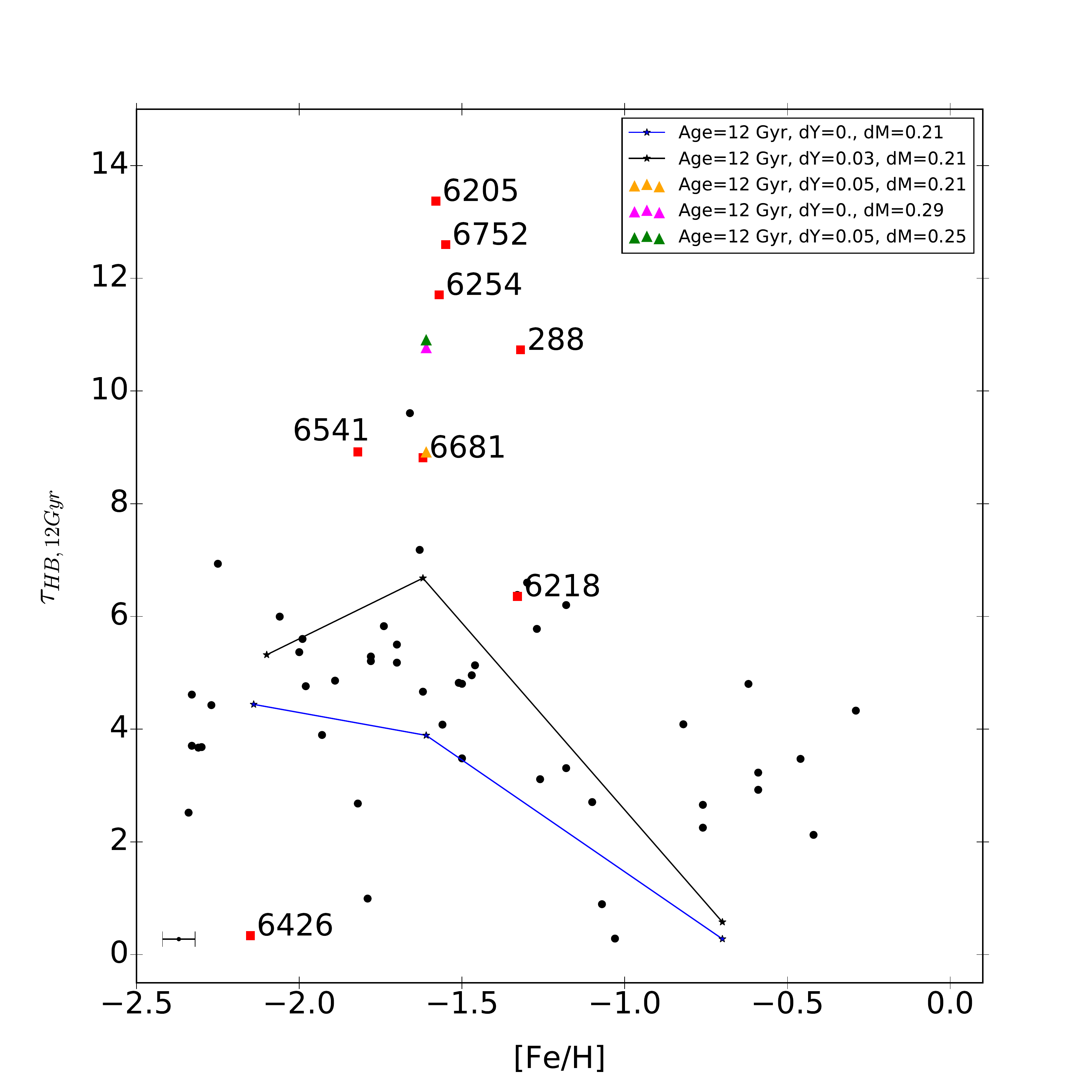}
\caption{The new HB morphology index, $\tau_{HB,12~Gyr}$, as a 
function of cluster iron abundance \citep{carretta}. 
The two lines display synthetic HB models at fixed cluster 
age (12~Gyr), but either with a canonical helium content ($\partial Y= 0$, blue line) 
or with an internal spread in He of $\partial Y$ = 0.03 (black line).
The coloured triangles identify three models for [Fe/H]=-1.6 for three 
different combinations of $\partial Y$ and $\Delta M$ (see legend).  
Red squares mark the eight second parameter clusters. 
The error bar on the left corner gives the 0.1 dex error on the 
metal content.}
\label{fig:TAUFEHsynth}
\end{figure}
%________________________________ 
Figure~\ref{fig:TAUFEHsynth} shows the comparison between the 
$\tau_{HB,12~Gyr}$ morphology index as a function of cluster iron abundance,
 and synthetic horizontal branch models.
The blue line shows the SHB model at constant helium 
content ($\partial Y= 0$), while the black one is the synthetic model constructed assuming 
an internal spread in He content of $\partial Y$ = 0.03. This value was 
adopted according to the recent estimates provided by \citet{milone18}. 

In the metal-rich regime 
([Fe/H]$\ge$-0.7) the predicted $\tau_{HB}$ weakly changes with 
the spread in helium content. This is a consequence 
of the fact that SHB models predict an extreme red HB morphology, 
minimally affected by the intrinsic parameters we are taking into account.
In the metal-intermediate regime ([Fe/H]=-1.62) the $\tau_{HB}$ value predicted
assuming an internal spread $\partial Y$ is characterized by a spike. This is due to 
the fact that at this metallicity values, a small mass variation causes significant changes
in the SHB colour.

To explain the existence of our second parameter clusters, the 
triangles in Fig.~\ref{fig:TAUFEHsynth} identify a further three different 
synthetic models for [Fe/H]=-1.6, the mean metallicity value of our second 
parameter clusters, at the fixed age of 12 Gyr. 
The yellow triangle shows the synthetic $\tau_{HB}$ considering a higher 
spread in He compared to the one estimated by \citet{milone18} 
($\partial Y = 0.05$) at the mass loss value we used for the blue and 
black models ($\Delta M=0.21$); the magenta triangle identifies the model for 
$\partial Y = 0$ and a mass loss of $\Delta M=0.29$ (higher of 0.08 $\msun$ than 
the standard one); finally the green triangle identifies the synthetic $\tau_{HB}$ 
value for the case with $\partial Y = 0.05$ and $\Delta M=0.25$. 

Therefore, Fig.~\ref{fig:TAUFEHsynth} shows that if we want to model the 
estimated values of $\tau_{HB,12 Gyr}$ we have three different possibilities. 
If we adopt the spread in He content estimated by \citet{milone18} (e.g.
$\partial Y=0.05$ for NGC~6205, $\partial Y=0.04$ for NGC~6752, 
$\partial Y=0.03$ for NGC~6254, $\partial Y=0.03$ for NGC~6681, considering
the second parameter cluster with $[Fe/H] \sim -1.6$), we need 
to increase the mass loss value to 0.29, which is a value greater by 0.08 $\msun$ 
than the one able to fit the bulk of the clusters. If we fix the mass loss 
to the one needed to model the bulk of the clusters ($\Delta M=0.21$), then we need 
to increase the spread in He content (from $\partial Y=0.03$ to, at least, 
$\partial Y=0.05$). Finally, we can explain the existence of the second 
parameter cluster also considering a higher mass loss together with a 
higher spread in He. We note that the adoption of a higher mass loss cannot be attributed to 
systematic errors in age estimations, since they should be on the order 
of 2-3 Gyr.
%_______________________________________________________________________________
\section{Conclusions}\label{sec:9}
We took advantage of a sample of 64 GGCs for which we have homogeneous and 
accurate $UBVRI$ ground photometry and $V$ ($F606W$), $I$ ($F814W$) ACS/$\hst$
 data \citep{sara,dotter11}, to introduce a new HB morphology index, named $\tau_{HB}$, 
 to investigate on a more quantitative basis the variation of  the HB morphology 
when moving from the metal-poor to the metal-rich regime. We define $\tau_{HB}$ as the ratio of the area below the cumulative number distribution 
in apparent magnitude ($A_{CND}(I)$) and in colour ($A_{CND}(\vmi)$) of the entire HB region.

Even though the estimate of the $\tau_{HB}$ index appears to be more complicated
 compared with HB morphology indices based either on star counts ($HBR'$, \citet{lee})
 or on specific evolutionary features (L1, L2, \citet{milone}), it offers several 
advantages. Indeed, we found that $\tau_{HB}$ is a factor of seven more sensitive than 
the classical  $HBR'$ index and more than one order of magnitude more sensitive
compared to L1 or L2 indices. Moreover, and even more importantly, the $\tau_{HB}$ 4index shows a 
linear trend over the entire metallicity range (-2.35$\leq$[Fe/H]$\leq$-0.12 ) covered by GGCs.  
Furthermore, the $\tau_{HB}$ index traces the HB luminosity function and it is independent 
of uncertainties affecting either the definition of different sub-groups
(blue, red, variables) or the position of specific evolutionary features (L1, L2). 

Moreover, to analyse the possible sensitivity of the HB morphology indices 
to the different contribution of inner and outer populations in GGCs, we estimated 
$HBR’$ and $\tau_{HB}$ considering just space-based and just ground-based data. 
Comparing them to the results obtained using the combined $\hst$ and ground-based 
observations, we found that $HBR’$ has on average lower differences 
($\sim 1.6 \%$, just ground-based, and $\sim 1.3 \%$, just space-based) 
than the ones attained by $\tau_{HB}$ ($\sim 3.6 \%$, just ground-based,  
and $\sim 2.9 \%$, just space-based). For $HBR’$ we observed major differences 
for the globulars in which the contribution of the red HB stars is mainly driven 
by $\hst$ observations. In $\tau_{HB}$ analysis we found higher differences for 
small and concentrated clusters, or for those clusters which are larger but 
located at higher distances. In all 
these cases most of the cluster stars are located inside the ACS FoV and so the 
space-based data give the higher contribution to the global $\tau_{HB}$.
In general our data are self-consistent and reliable and $\tau_{HB}$ higher 
relative differences could mean that it is sensitive to the different contribution 
given by inner and outer populations observed in GGCs.
 
To quantify the sensitivity of the $\tau_{HB}$ index on intrinsic stellar parameters, 
we investigated its dependence on cluster global properties 
(metallicity, absolute age, spread in He content). 
The main results of our analysis are the following:

\textit{Anti-correlation with cluster metallicity} -- We found a quadratic 
anti-correlation between $\tau_{HB}$ and [Fe/H]. 
The majority of the metal-poor globulars ([Fe/H]$\le$-1.5) have $\tau_{HB}$ 
between $\sim$ 4 and $\sim$ 9, while the metal-intermediate ones (-1.5$\le$ [Fe/H] $\le$-1.0) have values 
between $\sim$ 2 and $\sim$ 5. On the other hand, the metal-rich clusters, with [Fe/H] $\ge$-1.0, 
attain $\tau_{HB}$ values smaller than two.  

\textit{Identification of second parameter clusters} --  We found a 
subsample of eight GGCs in the metal-poor and metal-intermediate regime (-1.82$\le$ [Fe/H] $\le$-1.32) which, at fixed metallicity, are characterized
 by $\tau_{HB}$ values that are, on average, at least a factor of two larger than 
canonical clusters. The outlier clusters do not display any peculiarity 
in the $HBR'$ metallicity plane. To investigate their HB morphology,
 we selected three of them that sample the metal-rich ([Fe/H]=-1.33, NGC~6218), 
metal-intermediate ([Fe/H]=-1.57, NGC~6254), and metal-poor ([Fe/H]=-1.82, NGC~6541) 
regimes. We compared their $I, \vmi$ CMDs to those of three 'regular' clusters 
with similar metallicities within the errors (NGC~362, NGC~6934, and NGC~6144).
For each cluster pair, we find similar $HBR'$ values but different $\tau_{HB}$ values, 
with differences even on the order of three. For these reasons we can associate 
these clusters in our sample with the so-called second parameter
clusters. 

\textit{Correlation with cluster age} -- We investigate the relation between our HB morphology
index $\tau_{HB}$ and the absolute cluster age. To exclude possible dependence on the particular 
age estimation, we used the different homogeneous evaluations from
 \citet{vand, leaman,salarisweiss}, finding a linear correlation 
between$\tau_{HB}$ and the absolute age. We found that, in general, when 
moving from red to blue HB morphology, the GGCs become older. 
The second parameter clusters selected according to the $\tau_{HB}$-metallicity plane
appear to be peculiar also in the  $\tau_{HB}$-absolute age plane.
In particular they attain cluster ages ranging from $\sim$ 11.5 to $\sim$ 13 Gyr.
Moreover, they seem to be characterized by bluer HB morphologies than the typical clusters.

\textit{Reductio ad unum} -- We limited the age impact on our analysis
by reducing the $\tau_{HB}$ values to the ones they would attain for an age of 12~Gyr.  
In contrast to what we originally found, we found a linear correlation between the corrected 
$\tau_{HB, 12~Gyr}$ values and [Fe/H]. However, the second parameter clusters are still located 
far from the best-fit linear relation of the plane.

\textit{Comparison with spread in helium content} -- We investigated our new 
HB morphology index in the context of internal helium content variation, 
supposed to be one of the main drivers of the HB morphology. We compared the 
residuals of the corrected $\tau_{HB, 12 Gyr}$ as a function of cluster 
metallicity with the internal spread in helium content, $\partial Y$, estimated
by \citet{milone18}, but we did not find a solid correlation.

\textit{Comparison with theory} -- We calculated a
novel set of synthetic horizontal branch models to investigate the impact on $\tau_{HB}$
of the spread in Helium content and mass loss along the branch, at the fixed age of 
12 Gyr. We found that we can fit the second parameter clusters
in our sample if: we fix the spread in He content to the one estimated by 
\citet{milone18} and we increase the mass loss value from $\Delta M= 0.21$ (the one able
to fit the bulk of the clusters) to $\Delta M=0.29$; we fix the mass loss to $\Delta M=0.21$ 
and we increase the spread in He content from $\partial Y=0.03$ \citep{milone18} 
to, at least, $\partial Y=0.05$; we consider a higher mass loss together 
with an higher spread in He content.

\textit{Nature versus nurture} -- It is not clear whether the outlier clusters display 
a bluer HB morphology because they are intrinsically different (nature) or 
because the HB morphology is tracing a specific dynamical status of the cluster (nurture).
%%TABLES%%%
\setcounter{table}{0}
\begin{table*}
\scriptsize
\caption{List of the Globular Clusters in the sample with parameters used in this work: J~2000 coordinates, data availability in ground-based and space-based bands, tidal radius in arcmin, $HBR$ index, Age in Gyr, colour excess E(B-V), [Fe/H], apparent visual distance modulus $\mu$.}
\label{table:gcsuni}      
%\centering 
\begin{tabular}{c c c c c c c c c c c c c c c c c}  % 17 columns 
\hline \hline
 ID    &  Name   &  Ra (J~2000)\tablefootmark{a}     &Dec (J~2000) \tablefootmark{a} & $U$ & $B $&$V$& $R$& $I$& $F606W$&$F814W$ & $r_t (\arcmin) $ \tablefootmark{a}& HBR\tablefootmark{a} & Age(Gyr)\tablefootmark{b} & E(B-V)\tablefootmark{c} & [Fe/H]\tablefootmark{d} & $\mu$\tablefootmark{b}\\
\hline
NGC 0104 & 47 Tuc & 00:24:05.67 & -72:04:52.6&\cmark&\cmark&\cmark&\cmark&\cmark&\cmark&\cmark &42.86&-0.99 & 11.75 $\pm$ 0.25 & 0.03& -0.76 $\pm$ 0.02& 13.25 \\

NGC 0288 & & 00:52:45.24 & -26:34:57.4 &\cmark&\cmark&\cmark&\cmark&\cmark&\cmark&\cmark&12.94& 0.98 & 11.50 $\pm$ 0.38 & 0.01&-1.32 $\pm$ 0.02 & 14.87 \\

NGC 0362 & & 01:03:14.26& -70:50:55.6 &\cmark&\cmark&\cmark&\cmark&\cmark&\cmark&\cmark &16.11&-0.87 & 10.75 $\pm$ 0.25 & 0.03&-1.30 $\pm$ 0.04 & 14.70 \\

NGC 1261& & 03:12:16.21&-55:12:58.4&\cmark&\cmark&\cmark&\cmark&\cmark&\cmark&\cmark&7.28&-0.71& 10.75 $\pm$ 0.25 & 0.01&-1.27 $\pm$ 0.08 & 16.02 \\

NGC 1851& & 05:14:06.76 & -40:02:47.6& \cmark&\cmark&\cmark&\cmark&\cmark&\cmark&\cmark&11.7&-0.36& 11.00 $\pm$ 0.38 & 0.04&-1.18 $\pm$ 0.08 &  15.31 \\

NGC 2298 & & 06:48:59.41 &-36:00:19.1 &\cmark&\cmark&\cmark&\cmark&\cmark&\cmark&\cmark& 6.48&0.93 & - & 0.22&-1.96 $\pm$ 0.04 & - \\

NGC 3201& & 10:17:36.82 &-46:24:44.9&\cmark&\cmark&\cmark&\cmark&\cmark&\cmark&\cmark & 28.45&0.08 &11.50 $\pm$ 0.38&0.26 & -1.51 $\pm$ 0.02 & 13.29 \\

NGC 4147 & & 12:10:06.30&+ 18:32:33.5&\cmark&\cmark&\cmark&\cmark&\cmark&\cmark&\cmark& 6.31&0.55& 12.25 $\pm$ 0.25 & 0.03&-1.78 $\pm$ 0.08 & 16.33 \\

NGC 4590 &M 68 &12:39:27.98 &-26:44:38.6&\cmark&\cmark&\cmark&\cmark&\cmark&\cmark&\cmark&30.34&0.17 & 12.00 $\pm$ 0.25 & 0.06&-2.27 $\pm$ 0.04 & 16.05 \\	

NGC 4833 & & 12:59:33.92 & -70:52:35.4&\cmark&\cmark&\cmark&\cmark&\cmark&\cmark&\cmark & 17.85& 0.93& 12.50 $\pm$ 0.50  & 0.33&-1.89 $\pm$ 0.05 &   - \\

NGC 5024 & M 53& 13:12:55.25&+18:10:05.4&\cmark&\cmark&\cmark&\cmark&\cmark&\cmark&\cmark& 21.75&0.81& 12.25 $\pm$ 0.25 & 0.03&-2.06 $\pm$ 0.09 & 16.31 \\	

NGC 5053 & & 13:16:27.09&+17:42:00.9&\cmark&\cmark&\cmark&\cmark&\cmark&\cmark&\cmark& 13.67&0.50 &12.25 $\pm$ 0.38 & 0.02&-2.30 $\pm$ 0.08 & 16.19 \\

NGC 5272& M 3& 13:42:11.62&+28:22:38.2&\cmark&\cmark&\cmark&\cmark&\cmark&\cmark&\cmark& 38.19&0.08 &11.75 $\pm$ 0.25 & 0.01 &-1.50 $\pm$ 0.05 & 14.99 \\

NGC 5286 & & 13:46:26.81&-51:22:27.3&\cmark&\cmark&\cmark&\cmark&\cmark&\cmark&\cmark& 8.36&0.80&12.25 $\pm$ 0.38 & 0.29&-1.70 $\pm$ 0.07 &  15.04 \\	

NGC 5466 & & 14:05:27.29&+28:32:04.0&\cmark&\cmark&\cmark&\cmark&\cmark&\cmark&\cmark& 34.24&0.58 &12.50 $\pm$ 0.25 & 0.02&-2.31 $\pm$ 0.09 & 16.09 \\

NGC 5904& M 5 & 15:18:33.22 & +02:04:51.3&\cmark&\cmark&\cmark&\cmark&\cmark&\cmark&\cmark& 28.4&0.31 &11.50 $\pm$ 0.25 & 0.04& -1.33 $\pm$ 0.02 & 14.26 \\	

NGC 5927 & & 15:28:00.69 & -50:40:22.9 & \cmark&\cmark&\cmark&\cmark&\cmark&\cmark&\cmark& 16.68&-1.00 &10.75 $\pm$ 0.38 & 0.51&-0.29 $\pm$ 0.07 & 14.20 \\

NGC 5986 & & 15:46:03.00 &-37:47:11.1 &\cmark&\cmark&\cmark&\cmark&\cmark&\cmark&\cmark&10.52& 0.97&12.25 $\pm$ 0.75 &0.34 &-1.63 $\pm$ 0.08 & - \\
	
NGC 6093&  M 80 & 16:17:02.41 &-22:58:33.9&	\cmark&\cmark&\cmark&\cmark&\cmark&\cmark&\cmark& 13.28 &0.93 &- & 0.21&-1.75 $\pm$ 0.08 &-\\

NGC 6101 & & 16:25:48.12 &-72:12:07.9&\cmark&\cmark&\cmark&\cmark&\cmark&\cmark&\cmark& 7.27&0.84 &12.25 $\pm$ 0.50 & 0.10&-1.98 $\pm$ 0.07 &-\\

NGC 6121 & M 4 & 16:23:35.22 & -26:31:32.7 &\cmark&\cmark&\cmark&\cmark&\cmark&\cmark&\cmark& 32.49 &-0.06 &11.50 $\pm$ 0.38 &0.50 &-1.18 $\pm$ 0.02 & 11.07 \\	

NGC 6144 & & 16:27:13.86& -26:01:24.6 & 	&\cmark&\cmark&\cmark&\cmark&\cmark&\cmark& 33.25 & 1.00&12.75 $\pm$ 0.50 & 0.71&-1.82 $\pm$ 0.05 & - \\

NGC 6171 & M 107& 16:32:31.86 & -13:03:13.6 & \cmark&\cmark&\cmark&\cmark&\cmark&\cmark&\cmark& 17.44 &-0.73 &12.00 $\pm$ 0.75 &0.45 &-1.03 $\pm$ 0.02 & 13.43 \\	

NGC 6205 & M 13 & 16:41:41.24 & +36:27:35.5 &\cmark &\cmark&\cmark&\cmark&\cmark&\cmark&\cmark& 25.18&0.97 &12.00 $\pm$ 0.38 & 0.02&-1.58 $\pm$ 0.04 & 14.39 \\

NGC 6218 & M 12 &16:47:14.18&-01:56:54.7&\cmark&\cmark&\cmark&\cmark&\cmark&\cmark&\cmark &17.6&0.97 &13.00 $\pm$ 0.50 & 0.17&-1.33 $\pm$ 0.02 & 13.52 \\

NGC 6254 &M 10 &16:57:09.05&-04:06:01.1&\cmark&\cmark&\cmark&\cmark&\cmark&\cmark&\cmark&21.48 &1.00 &11.75 $\pm$ 0.38 & 0.29&-1.57 $\pm$ 0.02 & - \\

NGC 6304 & & 17:14:32.25 &-29:27:43.3 &&&&&&\cmark&\cmark& 13.25&-1.00&11.25 $\pm$ 0.38 & 0.52&-0.37 $\pm$ 0.07 & 13.81 \\

NGC 6341 & M 92 & 17:17:07.39 & +43:08:09.4&\cmark&\cmark&\cmark&\cmark&\cmark&\cmark&\cmark &15.17&0.91 &12.75 $\pm$ 0.25 & 0.02&-2.35 $\pm$ 0.05 & 14.66 \\

NGC 6352 & & 17:26:29.11 & -48:25:19.8 & &\cmark&\cmark&&\cmark&\cmark&\cmark& 10.51 &-1.00&10.75 $\pm$ 0.38 & 0.35&-0.62 $\pm$ 0.05 & 13.36 \\

NGC 6362 & & 17:31:54.99 &-67:02:54.0 & 	&\cmark&\cmark&\cmark&\cmark&\cmark&\cmark& 16.67&-0.58&12.50 $\pm$ 0.25 & 0.07&-1.07 $\pm$ 0.05 & 14.36 \\

NGC 6366 & & 17:27:44.24 & -05:04:47.5 & 	\cmark&\cmark&\cmark&\cmark&\cmark&\cmark&\cmark& 15.2&-0.97 &11.00 $\pm$ 0.50 &0.75 &-0.59 $\pm$ 0.08 & 12.40 \\

NGC 6388 & & 17:36:17.23 & -44:44:07.8 & 	&\cmark&\cmark&&\cmark&\cmark&\cmark& 6.21 &- & - & 0.41&-0.45 $\pm$ 0.04 &-\\

NGC 6397 & & 17:40:42.09 & -53:40:27.6 & 	\cmark&\cmark&\cmark&\cmark&\cmark&\cmark&\cmark& 15.81 & 1.00 &13.00 $\pm$ 0.25 & 0.19&-1.99 $\pm$ 0.02 & -\\

NGC 6426 & & 17:44:54.65 &+03:10:12.5&	&&&&&\cmark&\cmark& 13.23&0.58 &- & 0.35&-2.15 &-\\	

NGC 6441 & & 17:50:13.06 & -37:03:05.2 &	\cmark&\cmark&\cmark&\cmark&\cmark&\cmark&\cmark& 8.0 &- &- & 0.61&-0.44 $\pm$ 0.07 &-\\

NGC 6496 & & 17:59:03.68 & -44:15:57.4 & \cmark&\cmark&\cmark&\cmark&\cmark&\cmark&\cmark& 5.27& -1.00 &10.75 $\pm$ 0.38 & 0.23&-0.46 $\pm$ 0.07 & 14.93 \\

NGC 6535 & & 18:03:50.51 & -00:17:51.5 &\cmark&\cmark&\cmark&&\cmark&\cmark&\cmark& 8.36 & 1.00 &12.75 $\pm$ 0.50 & 0.41&-1.79 $\pm$ 0.07 & 13.88 \\

NGC 6541 & & 18:08:02.36&-43:42:53.6&\cmark&\cmark&\cmark&&\cmark&\cmark&\cmark& 29.6&1.00&12.50 $\pm$ 0.50 & 0.16&-1.82 $\pm$ 0.08 & - \\	

NGC 6584 & & 18:18:37.60 & -52:12:56.8 &&\cmark&\cmark&\cmark&\cmark&\cmark&\cmark& 9.37&-0.15 &11.75 $\pm$ 0.25 & 0.11&-1.50 $\pm$ 0.09 & 15.54\\

NGC 6624 & & 18:23:40.51&-30:21:39.7&&&&&&\cmark&\cmark& 20.55&-1.00 &11.25 $\pm$ 0.50 & 0.26&-0.42 $\pm$ 0.07 & 14.47 \\

NGC 6637 & M 69 & 18:31:23.10 & -32:20:53.1 && \cmark&\cmark&&\cmark&\cmark&\cmark&  8.35 & -1.00 &11.00 $\pm$ 0.38 & 0.17&-0.59 $\pm$ 0.07 & 14.70\\

NGC 6652 & & 18:35:45.63 & -32:59:26.6 & 	&&\cmark&&\cmark&\cmark&\cmark &  4.48 & -1.00&11.25 $\pm$ 0.25 & 0.11&-0.76 $\pm$ 0.14 & 14.90 \\

NGC 6656 & M 22& 18:36:23.94 & -23:54:17.1 & 	\cmark&\cmark&\cmark&\cmark&\cmark&\cmark&\cmark&  28.97 &0.91 &12.50 $\pm$ 0.50 & 0.33&-1.70 $\pm$ 0.08 & -\\

NGC 6681 & M 70 & 18:43:12.76 & -32:17:31.6 & \cmark&\cmark&\cmark&\cmark&\cmark&\cmark&\cmark& 7.91 & 0.96&12.75 $\pm$ 0.38 & 0.11&-1.62 $\pm$ 0.08 & 14.78 \\

%NGC 6715 & M 54 & 18:55:03.33 & -30:28:47.5 & &\cmark&\cmark&\cmark&\cmark&\cmark&\cmark &  7.47 &0.75 &11.75 $\pm$ 0.50 & 0.15&-1.44 $\pm$ 0.07 & 17.08 \\

NGC 6717 &Pal 9 & 18:55:06.04 & -22:42:05.3 & 	&\cmark&\cmark&\cmark&\cmark&\cmark&\cmark&  9.87 &0.98&12.50 $\pm$ 0.50 & 0.25&-1.26 $\pm$ 0.07 & 14.17 \\

NGC 6723& & 18:59:33.15 & -36:37:56.1&	&\cmark&\cmark&&\cmark&\cmark&\cmark& 10.51&-0.08 &12.25 $\pm$ 0.25 & 0.16&-1.10 $\pm$ 0.07 & 14.23 \\

NGC 6752 & & 19:10:52.11 & -59:59:04.4 & 	\cmark&\cmark&\cmark&\cmark&\cmark&\cmark&\cmark& 55.34 &1.00 &12.50 $\pm$ 0.25 & 0.06&-1.55 $\pm$ 0.01 & - \\

NGC 6779 & M 56 & 19:16:35.57 & +30:11:00.5 & \cmark&\cmark&\cmark&\cmark&\cmark&\cmark&\cmark& 8.56&0.98 &12.75 $\pm$ 0.50 & 0.25&-2.00 $\pm$ 0.09 & -\\ 	

NGC 6809 & M 55 & 19:39:59.71 & -30:57:53.1 &	\cmark&\cmark&\cmark&\cmark&\cmark&\cmark&\cmark& 16.28& 0.87 &13.00 $\pm$ 0.25 & 0.09&-1.93 $\pm$ 0.02 & 13.62\\

NGC 6838 & M 71 & 19:53:46.49 & +18:46:45.1 & \cmark&\cmark&\cmark&\cmark&\cmark&\cmark&\cmark& 8.96&-1.00&11.00 $\pm$ 0.38 & 0.33&-0.82 $\pm$ 0.02 & 12.67 \\

NGC 6934 & & 20:34:11.37 & +07:24:16.1 & \cmark&\cmark&\cmark&\cmark&\cmark&\cmark&\cmark&  8.37 & 0.25&11.75 $\pm$ 0.25 & 0.11&-1.56 $\pm$ 0.09 & 16.05 \\

NGC 6981 & M 72& 20:53:27.70 & -12:32:14.3 &\cmark&\cmark&\cmark&\cmark&\cmark&\cmark&\cmark& 9.15&0.14 &11.50 $\pm$ 0.25 & 0.06&-1.48 $\pm$ 0.07 & 16.05 \\

NGC 7006 & & 21:01:29.38 & +16:11:14.4&\cmark&\cmark&\cmark&\cmark&\cmark&\cmark&\cmark & 6.34&-0.28 &	- & 0.08&-1.46 $\pm$ 0.06 & - \\

NGC 7078 & M 15 & 21:29:58.33 &+12:10:01.2&\cmark&\cmark&\cmark&\cmark&\cmark&\cmark&\cmark& 21.5&0.67 & 12.75 $\pm$ 0.25 & 0.11&-2.33 $\pm$ 0.02 & 15.03 \\

NGC 7089 & M 2& 21:33:27.02 & -00:49:23.7 &\cmark&\cmark&\cmark&\cmark&\cmark&\cmark&\cmark& 21.45& 0.96 & 11.50 $\pm$ 0.25 & 0.04&-1.66 $\pm$ 0.07 & 15.41 \\	

NGC 7099 & M 30 & 21:40:22.12&-23:10:47.5&\cmark&\cmark&\cmark&\cmark&\cmark&\cmark&\cmark& 18.34 & 0.89 &  13.00 $\pm$ 0.25 & 0.05&-2.33 $\pm$ 0.02 & 14.60 \\	

ARP 2 &  & 19:28:44.11 & -30:21:20.3 &\cmark& \cmark&\cmark&&\cmark&\cmark&\cmark & 12.65&0.86 & 12.00 $\pm$ 0.38 & 0.11&-1.74 $\pm$ 0.08 & 17.25 \\

IC 4499 & &15:00:18.45&-82:12:49.3  &\cmark&\cmark&\cmark&\cmark&\cmark&\cmark&\cmark&12.35&0.11 &	- & 0.22&-1.62 $\pm$ 0.09 &	-\\		

Lynga 7 & &16:11:03.65  &-55:19:04.0 & &&&&&\cmark&\cmark&-& -1.00 & -& 1.06&- & -\\

%Pal 1 & & 03:33:20.04 & +79:34:51.8 &&&\cmark&&\cmark&\cmark&\cmark& 8.96&-1.00 & - & 0.20&-0.51 $\pm$ 0.09 &-\\

Pal 2 & & 04:46:05.91 & +31:22:53.4 & &\cmark&\cmark&\cmark&\cmark&\cmark&\cmark  & 6.76&-& - & 1.21&-1.29 $\pm$ 0.09 & -\\

%Pal 12 & & 21:46:38.84 &-21:15:09.4&	&\cmark&\cmark&\cmark&\cmark&\cmark&\cmark & 17.42&-1.00 & 9.0 $\pm$ 0.38 & 0.04&-0.81 $\pm$ 0.08 & 16.36 \\

Rup 106 & &12:38:40.2  &-51:09:01 & 	&\cmark&\cmark&\cmark&\cmark&\cmark&\cmark&5.03&-0.82 &	 - & 0.17&-1.78 $\pm$ 0.08 &-\\

Terzan 7 & & 19:17:43.92 & -34:39:27.8&	&\cmark&\cmark&\cmark&\cmark&\cmark&\cmark&  7.27 &-1.00 & - & 0.09&-0.12 $\pm$ 0.08 & -\\

Terzan 8 & & 19:41:44.41 & -33:59:58.1 & \cmark&\cmark&\cmark&\cmark&\cmark&\cmark&\cmark&  4.0 &1.00& 13.00 $\pm$ 0.38  & 0.15&-& 17.14 \\	
\hline                                 
\end{tabular}
\tablefoot{
\tablefoottext{a}{\cite{harris};}
\tablefoottext{b}{\cite{vand};}
\tablefoottext{c}{\cite{dutra}; }
\tablefoottext{d}{\cite{carretta}.}
}
\end{table*}
%% TABLE 2A %%
\begin{table*}
\scriptsize
\caption{Observational values of $HBR'$, blue (B) and red (R) HB star counts. L1, L2 indices, the spread in helium content $\partial Y$.  V(HB) and $A_{CND}(V-I)$ we used to estimate the new parameter $\tau_{HB}$.}   
\label{table:hbr}      
\centering 
\begin{tabular}{c c c c c c c c c c c c}  % 12 columns 
\hline\hline       
ID   & Name & [Fe/H]  &  B    &  R    &    $HBR'$ & L1\tablefootmark{a}& L2\tablefootmark{a} & $\partial Y $ \tablefootmark{b} &V(HB)\tablefootmark{c}  & $A_{CND}(V-I)$ &$\tau_{HB}$\\
\hline   
NGC~0104 & 47~Tuc  &  -0.76  &  0 	& 1336  	& 1.01   $\pm$  0.04 & 0.078 $\pm$ 0.005 & 0.068 $\pm$ 0.006 & 0.049 $\pm$ 0.005& 14.06 & 137.9 $\pm$ 11.7  & 1.58 $\pm$ 0.01	\\
NGC~0288   & & -1.32  & 130   & 	 0      &  2.98   $\pm$  0.12 & 0.534 $\pm$ 0.086 & 0.337 $\pm$ 0.086 & 0.016 $\pm$0.012 &15.44 & 47.8 $\pm$ 6.9 &9.39 $\pm$ 0.37	\\
NGC~0362   & & -1.30  &  12    & 282      & 1.24   $\pm$  0.07 & 0.086 $\pm$ 0.005 & 0.608 $\pm$ 0.064 &0.026 $\pm$0.008 &15.44 & 126.1 $\pm$ 11.2& 3.24 $\pm$ 0.01\\
NGC~1261  & & -1.27  &  35   &  201	    & 1.36   $\pm$  0.06 & 0.088 $\pm$ 0.005 & 0.644 $\pm$ 0.038 & 0.019 $\pm$ 0.007 &16.70 &114.0 $\pm$ 10.7 & 2.42 $\pm$ 0.02	\\
NGC~1851  & & -1.18  & 137   &  156     & 1.95   $\pm$  0.01 & 0.098 $\pm$ 0.004 & 0.679 $\pm$ 0.010  & 0.025 $\pm$ 0.006 & 16.09 & 90.9 $\pm$ 9.5 & 3.51 $\pm$ 0.06	\\
NGC~2298  & & -1.96	&  53	&	 1		&  2.90   $\pm$  0.17 & 0.486 $\pm$ 0.020 & 0.267 $\pm$ 0.023 &0.011 $\pm$ 0.012&16.11 & 57.7 $\pm$ 7.6 & 5.97 $\pm$ 0.17   \\
NGC~3201  & & -1.51	&  90	&  103		& 1.96   $\pm$  0.01 & 0.106 $\pm$ 0.015 & 0.649 $\pm$ 0.022 & 0.028 $\pm$ 0.032 & 14.76 & 88.7 $\pm$ 9.4 & 2.80 $\pm$ 0.05\\
NGC~4147  & & -1.78  &  61    &   2      &  2.76   $\pm$  0.13 & 0.271 $\pm$ 0.027 & 0.476 $\pm$ 0.029 & - &17.02 & 61.7 $\pm$ 7.9 & 5.96 $\pm$ 0.12  	\\
NGC~4590 & M~68  &  -2.27  &  61    &  1    	&  2.58   $\pm$  0.09 & 0.205 $\pm$ 0.029 & 0.524 $\pm$ 0.030 & 0.012 $\pm$ 0.009 & 15.68 & 74.8 $\pm$ 8.6 & 4.43 $\pm$0.07\\
NGC~4833  & & -1.89	& 280	&   8		&  2.88   $\pm$  0.07  & 0.287 $\pm$ 0.037 & 0.538 $\pm$ 0.037 & 0.051 $\pm$ 0.009 & 15.60 & 55.2 $\pm$ 7.4 & 6.21 $\pm$ 0.28\\ 
NGC~5024 & M~53 &  -2.06  & 495   &    0      &  2.89  $\pm$  0.06& 0.158 $\pm$ 0.035 & 0.602 $\pm$ 0.036 & 0.04 $\pm$ 0.008 &16.81 & 57.9 $\pm$ 7.6  & 6.67 $\pm$ 0.16	\\
NGC~5053 & &  -2.30	&  34	&	 8		&  2.46   $\pm$  0.12 & 0.223 $\pm$ 0.090 & 0.439 $\pm$ 0.090 & 0.004 $\pm$ 0.35 &16.69 & 74.2 $\pm$ 8.6 & 4.35 $\pm$ 0.11  \\
NGC~5272 & M~3 &  -1.50  & 213  &  392		&  2.21   $\pm$  0.02 & 0.150 $\pm$ 0.016 & 0.613 $\pm$ 0.018 & 0.041 $\pm$ 0.009 &15.64 & 81.5 $\pm$ 9.0& 4.13 $\pm$ 0.05	\\ 
NGC~5286 & &  -1.70	& 548	&   36		&  2.80   $\pm$  0.05 & 0.213 $\pm$ 0.034 & 0.670 $\pm$ 0.035 & 0.044 $\pm$ 0.004 & 16.63 & 53.7 $\pm$ 7.3 & 6.17 $\pm$ 0.30  \\
NGC~5466 & &	 -2.31	&  72	&	 5		&  2.62   $\pm$  0.11 & 0.225 $\pm$ 0.062 & 0.457 $\pm$ 0.063 & 0.007 $\pm$ 0.024 & 16.52 & 68.7 $\pm$ 8.3 & 5.02 $\pm$ 0.10  \\
NGC~5904 & M~5  &  -1.33  & 98   &  155		& 2.42   $\pm$  0.03 & 0.150 $\pm$ 0.012 & 0.681 $\pm$ 0.014 & 0.037 $\pm$ 0.007 &15.07 & 68.9 $\pm$ 8.3 & 5.04 $\pm$ 0.14	\\
NGC~5927 & &  -0.29	&	0	&  447		& 1.00   $\pm$  0.07 & 0.043 $\pm$ 0.003 & 0.062 $\pm$ 0.004  & 0.055 $\pm$ 0.015 & 16.55 & 135.2 $\pm$ 11.6 & 0.97 $\pm$ 0.01   \\
NGC~5986 & &  -1.63 	& 459	&    26		& 2.88 $\pm$  0.06 & 0.460 $\pm$ 0.052 & 0.443 $\pm$ 0.053 & 0.031 $\pm$ 0.012 & 16.52 & 46.1  $\pm$ 6.8& 7.85 $\pm$ 0.46  \\  
NGC~6093 & M~80 &  -1.75 	& 332	&    16		&  2.87  $\pm$  0.07 & 0.464 $\pm$ 0.059 & 0.447 $\pm$ 0.062 & 0.027 $\pm$ 0.012 &16.10 &53.8 $\pm$ 7.3& 7.86 $\pm$ 0.36    \\
NGC~6101 & &  -1.98 	& 169	&	 17		&  2.87   $\pm$  0.09 & 0.485 $\pm$ 0.025 & 0.223 $\pm$ 0.025 & 0.017 $\pm$ 0.011 & 16.60 & 62.4 $\pm$ 7.9 & 5.43 $\pm$ 0.15 \\
NGC~6121 & M~4 &  -1.18	&  51	&   55		& 1.98   $\pm$  0.01 & 0.120 $\pm$ 0.020 & 0.569 $\pm$ 0.020& 0.014 $\pm$ 0.006 &13.45 & 86.7 $\pm$ 9.3 & 1.96 $\pm$ 0.06   \\
NGC~6144 & &  -1.82	&  62 	&   0	 	& 3.00  $\pm$  0.18 & 0.533 $\pm$ 0.024 & 0.229 $\pm$ 0.023 & 0.017 $\pm$ 0.013 &16.40 & 28.6 $\pm$ 5.3 & 4.98 $\pm$ 0.88  \\ %new
NGC~6171 & M~107 & -1.03	& 12 	&  89 	& 1.37 $\pm$  0.09 & 0.100 $\pm$ 0.014 & 0.513 $\pm$ 0.074 & 0.024 $\pm$ 0.014 & 15.00 & 116.2 $\pm$ 10.8 & 0.28 $\pm$ 0.03   \\ %new
NGC~6205 & M~13 &  -1.58  & 705   &    0      &  2.99   $\pm$  0.05 & 0.527 $\pm$ 0.013 & 0.441 $\pm$ 0.012 & 0.052 $\pm$ 0.004 &14.90 & 41.0 $\pm$ 6.4 & 13.37 $\pm$ 0.72	\\
NGC~6218 & M~12 &  -1.33	& 188	&	 0		&  3.00   $\pm$  0.10 & 0.561 $\pm$ 0.034 & 0.299 $\pm$ 0.035 & 0.011 $\pm$ 0.011 & 14.60 & 49.8 $\pm$ 7.1 & 9.05 $\pm$ 0.32 \\	
NGC~6254 & M~10 &  -1.57	& 257	&	 0		&  3.00   $\pm$  0.09 & 0.588 $\pm$ 0.032 & 0.260 $\pm$ 0.033 & 0.029 $\pm$ 0.011 &14.65 & 40.6 $\pm$ 6.4 & 11.03 $\pm$ 0.66  \\
NGC~6304  & &  -0.37	& 0&  170	& 1.03 $\pm$  0.10 &  0.062 $\pm$ 0.007 & 0.060 $\pm$ 0.004 & 0.025 $\pm$ 0.006 &15.60 &  0 & 0  \\ %new
NGC~6341 & M~92 &  -2.25  & 340   &   6    	&  2.92   $\pm$  0.07 & 0.261 $\pm$ 0.075 & 0.542 $\pm$ 0.075  & 0.039 $\pm$ 0.006& 15.10 & 47.4 $\pm$ 6.9 & 8.95 $\pm$ 0.30  \\
NGC~6352 & &  -0.62  &   0	&   96   	& 1.00   $\pm$  0.14 & 0.072 $\pm$ 0.007 & 0.056 $\pm$ 0.003 & 0.027 $\pm$ 0.006 &15.13 & 130.6 $\pm$ 11.4 & 1.44 $\pm$ 0.01  \\
NGC~6362 & &  -1.07  &  56   &  135   	& 1.65   $\pm$  0.07 & 0.122 $\pm$ 0.004 & 0.621 $\pm$ 0.039  & 0.004 $\pm$ 0.011 &15.33 &115.1 $\pm$ 10.7 & 2.24 $\pm$ 0.02 \\
NGC~6366 & &  -0.59	&	0	&   39		& 1.02	  $\pm$  0.22  & 0.076 $\pm$ 0.005 & 0.058 $\pm$ 0.018 & 0.011 $\pm$ 0.011 & 15.65 & 136.9 $\pm$ 11.7 & 0.23 $\pm$ 0.01  \\
NGC~6388 & &  -0.45	& 267	& 1549		& 1.30  $\pm$  0.03 & 0.057 $\pm$ 0.004 & 0.836 $\pm$ 0.008  & 0.067 $\pm$ 0.009 &16.85 & 119.5 $\pm$ 10.9 & 1.88 $\pm$ 0.03  \\
NGC~6397 & &-1.99	&120	&	 0		&  2.98   $\pm$	 0.13 & 0.534 $\pm$ 0.023 & 0.232 $\pm$ 0.030 & 0.008 $\pm$ 0.011 & 12.87 & 51.3 $\pm$ 7.1& 8.29 $\pm$ 0.17  \\
NGC~6426 & &  -2.15	& 46 	&   11 	&2.49 $\pm$  0.10 & 0.178 $\pm$ 0.018 & 0.519 $\pm$ 0.023 & 0.021 $\pm$ 0.006 & 18.16 & 75.8 $\pm$ 8.7 & 1.01 $\pm$ 0.16  \\ % metallicità da Harris %new
NGC~6441 & &  -0.44	& 243	& 1769		& 1.26   $\pm$  0.02 & 0.048 $\pm$ 0.003 & 0.904 $\pm$ 0.024 & 0.081 $\pm$ 0.022 & 17.51 & 113.9 $\pm$ 10.7 & 1.55 $\pm$ 0.03    \\ 
NGC~6496 & & -0.46	& 0	&   80 	&1.00 $\pm$  0.16 & 0.074 $\pm$ 0.011 & 0.056 $\pm$ 0.005  & 0.021 $\pm$ 0.006 & 16.00 & 143.2 $\pm$ 12.0 & 0.11 $\pm$ 0.02 \\ %new
NGC~6535 & &  -1.79	& 24	&   0 	&3.00 $\pm$  0.29 & 0.510 $\pm$ 0.026 & 0.271 $\pm$ 0.031   & 0.003 $\pm$ 0.022 & 15.75 & 53.0 $\pm$ 7.2 & 3.01 $\pm$ 0.33 \\ %new
NGC~6541 & &  -1.82  & 411   &    0      &  2.99  $\pm$  0.07 & 0.563 $\pm$ 0.026 & 0.347 $\pm$ 0.033 & 0.045 $\pm$ 0.006 & 15.35 & 41.7 $\pm$ 6.4 & 10.26 $\pm$ 0.46 	\\
NGC~6584 & &  -1.50	&  25	&   38	 	& 1.90  $\pm$  0.03 & 0.102 $\pm$ 0.012 & 0.558 $\pm$ 0.026 & 0.015 $\pm$ 0.011 & 16.53 & 94.4 $\pm$ 9.7 & 2.81 $\pm$ 0.05   \\
NGC~6624 & &  -0.42	& 0&   192 	&1.00 $\pm$  0.10 & 0.077 $\pm$ 0.006 & 0.085 $\pm$ 0.006  & 0.022 $\pm$ 0.003 & 15.60 & 141.8 $\pm$ 11.9 & 0.11 $\pm$ 0.02  \\ %new
NGC~6637 & &  -0.59	& 0&   256	&1.00 $\pm$  0.09 & 0.078 $\pm$ 0.004 & 0.065 $\pm$ 0.005  & 0.011 $\pm$ 0.005 & 15.34 & 137.8 $\pm$ 11.7 & 0.54 $\pm$ 0.02 \\ %new
NGC~6652 & &  -0.76	& 0&   26	&1.00 $\pm$  0.28 & 0.073 $\pm$ 0.011 & 0.080 $\pm$ 0.012  &0.017 $\pm$ 0.011 & 15.40 & 139.8 $\pm$ 11.8 & 0.64 $\pm$ 0.01  \\ %new
NGC~6656 & M~22 &  -1.70	& 512	&   13		&  2.94   $\pm$  0.06 & 0.336 $\pm$ 0.088 & 0.577 $\pm$ 0.087 & 0.041 $\pm$ 0.012 & 14.15 & 55.1 $\pm$ 7.4 & 6.53 $\pm$ 0.31 \\
NGC~6681 &  M~70 & -1.62 & 147 & 0 & 2.97 $\pm$ 0.11 & 0.558 $\pm$ 0.046 & 0.334 $\pm$ 0.045 & 0.029 $\pm$ 0.015 & 15.55 & 45.1 $\pm$ 6.7 &10.21 $\pm$ 0.37 \\ %new 
NGC~6717& &  -1.26	& 30&   0	&3.00 $\pm$  0.25 & 0.495 $\pm$ 0.032 & 0.310 $\pm$ 0.033 & 0.003 $\pm$ 0.009 & 15.55 & 52.3 $\pm$ 7.2 & 4.46 $\pm$ 0.24 \\ %new
NGC~6723 & &  -1.10  & 106   &  152	& 1.85 $\pm$  0.02 & 0.127 $\pm$ 0.007 & 0.704 $\pm$ 0.010  & 0.024 $\pm$ 0.007 &15.48 & 83.4 $\pm$ 9.1 &3.38 $\pm$ 0.08 	\\
NGC~6752 & &  -1.55  & 330   &    0      &  3.00   $\pm$  0.08 & 0.378 $\pm$ 0.024 & 0.578 $\pm$ 0.025 & 0.042 $\pm$ 0.004 &13.70 & 39.4 $\pm$ 6.3 & 13.94 $\pm$ 0.78 	\\
NGC~6779& M~56 & -2.00	& 177&   0	&3.00 $\pm$  0.10 & 0.508 $\pm$ 0.031 & 0.284 $\pm$ 0.030   & 0.031 $\pm$ 0.008 & 16.18 & 47.1 $\pm$ 6.9 & 7.38 $\pm$ 0.25 \\ %new
NGC~6809 & M~55 &  -1.93  & 244  &   2      &  2.93  $\pm$  0.08 & 0.476 $\pm$ 0.055 & 0.313 $\pm$ 0.053  & 0.026 $\pm$ 0.015 & 14.40 & 58.3 $\pm$ 7.6 & 6.59 $\pm$ 0.21	\\
NGC~6838 & M~71 &  -0.82	&   0	&   70		& 1.00   $\pm$  0.17 & 0.084 $\pm$ 0.012 & 0.057 $\pm$ 0.010 & 0.024 $\pm$ 0.010 & 14.48 & 126.5 $\pm$ 11.2 & 1.40 $\pm$ 0.01  \\
NGC~6934 & &  -1.56	& 103	&   71		&  2.13   $\pm$  0.02 & 0.097 $\pm$ 0.013 & 0.678 $\pm$ 0.016  & 0.018 $\pm$ 0.004 & 16.86 & 83.7 $\pm$ 9.1 & 3.41 $\pm$ 0.08  \\
NGC~6981 & M~72 &  -1.48  &  37  &   27  	& 2.14   $\pm$  0.03 & 0.142 $\pm$ 0.016 & 0.570 $\pm$ 0.019  & 0.017 $\pm$ 0.006 &16.90 & 84.9 $\pm$ 9.2 & 3.61 $\pm$ 0.06	\\
NGC~7006 & &  -1.46	&  75 	&   93		& 1.92   $\pm$  0.02 & 0.123 $\pm$ 0.016 & 0.581 $\pm$ 0.018 & - &18.80 & 92.5 $\pm$ 9.6 & 3.12 $\pm$ 0.05   \\
NGC~7078 & M~15 &  -2.33  &  554  &   56 &  2.64   $\pm$  0.04 & 0.174 $\pm$ 0.011 & 0.713 $\pm$ 0.019  & 0.069 $\pm$ 0.006 &15.83 & 58.3 $\pm$ 7.6 & 6.63 $\pm$ 0.24 	\\
NGC~7089 & M~2  &  -1.66  &  896 &   14    	&  2.92   $\pm$  0.04 & 0.150 $\pm$ 0.035 & 0.790 $\pm$ 0.037 & 0.052 $\pm$ 0.009 &16.05 & 50.8 $\pm$ 7.1 & 8.23 $\pm$ 0.47	\\
NGC~7099 & M~30 &  -2.33	& 175 	&   6		&  2.90   $\pm$  0.10 & 0.462 $\pm$ 0.103 & 0.261 $\pm$ 0.103 & 0.022 $\pm$ 0.010 &15.10 & 57.7 $\pm$ 7.6 & 6.40 $\pm$ 0.20  \\
ARP~2    & &  -1.74	&  22	&    0		&  2.86   $\pm$  0.10 & 0.491 $\pm$ 0.021 & 0.184 $\pm$ 0.021  & - & 18.13 & 64.7 $\pm$ 8.0 & 5.83 $\pm$ 0.05 \\
IC~4499 & &  -1.62  &  81  &   108   	& 1.90   $\pm$  0.02 & 0.113 $\pm$ 0.026 & 0.508 $\pm$ 0.041 & 0.017 $\pm$ 0.008 & 17.65 & 94.7 $\pm$ 9.7 & 2.65 $\pm$ 0.04\\
Lynga~7 & &  -0.67	& 0&  90	&1.00 $\pm$  0.15 & 0.055 $\pm$ 0.050 & 0.093 $\pm$ 0.009 &-& 16.70 & 0 & 0  \\ % metallicità da Harris 
Pal~2 & & -1.29	& 194&   0	&3.00 $\pm$  0.10 & - & - &-&20.85 & 88.2 $\pm$ 9.4 & 0.11 $\pm$ 0.08   \\ %new
Rup~106    &	& -1.78	&  42	&    0		&  2.78	  $\pm$  0.16 & 0.135 $\pm$ 0.012 & 0.210 $\pm$ 0.032  & - & 17.80 & 119.2 $\pm$ 10.9 & 1.84 $\pm$ 0.02   \\
Terzan~7 & & -0.12	& 0&   38	&1.00 $\pm$  0.23 & 0.057 $\pm$ 0.007 & 0.032 $\pm$ 0.008 & -&17.50 & 133.7 $\pm$ 11.6 & 0.53 $\pm$ 0.01   \\ %new
Terzan~8 & & -2.34	& 49&   0	&2.92 $\pm$  0.19 & 0.500 $\pm$ 0.050 & 0.223 $\pm$ 0.050 & -&17.95 & 55.9 $\pm$ 7.5 &  5.21 $\pm$ 0.17 \\ %new
\hline                  
\end{tabular}
\tablefoot{
\tablefoottext{a}{\cite{harris}}
\tablefoottext{b}{\cite{milone18}}
\tablefoottext{c}{\cite{milone}}
\tablefoottext{d}{Estimated by the authors}
}
\end{table*}

%_______________________________________________________________________________
\begin{acknowledgements}
This work has made use of data from \textit{An ACS Survey 
of Galactic Globular Clusters} programme in
the context of the Hubble Space Telescope Treasury project.
This investigation was partially supported by PRIN-INAF 2016 ACDC(P.I.:P. Caraveo). M.M. was supported by the Spanish Ministry of Economy
and Competitiveness (MINECO) under the grant AYA2017-89076-P.
\end{acknowledgements}
%_______________________________________________________________________________
%BIBLIOGRAPHY
\bibliographystyle{aa} % style aa.bst
\bibliography{35995corr_m} % your references Yourfile.bib
\begin{appendix}
%_______________________________________________________________________________

\section{Peculiar globular clusters}\label{app:A}
The classical HB morphology index $(HBR'$) estimations for NGC~6304, NGC~6426, 
NGC~6624, Lynga~7, and Palomar~2 and the $\tau_{HB}$ indices for NGC~6426, 
NGC~6624, and NGC~6652 were derived using $I$, $V$-bands from ACS-$\hst$
because we lack ground-based catalogues. For these globulars we considered 
all the stars in the $\hst$ field and within the tidal radius in ground-based telescope fields.

Due to the high field contamination in ground observations, for the clusters NGC~6388 
and NGC~6441, we used only ACS data. Despite this, the sample is statistically 
good enough to analyse their HB morphology, even if the ACS FoV does 
not cover the entire extent of these two clusters.

Owing to the low number of HB stars in Lynga~7 and NGC~6304, we obtain 
$\tau_{HB}$=0 and so, these clusters have not been included in the analysis 
of our new index.
Finally, we note that ground-based data for NGC~104, NGC~5272, NGC~5466, NGC~5927, NGC~6362, 
NGC~6397, and NGC~6752 do not reach the tidal radius.
%_______________________________________________________________________________
\section{Notes on individual outliers}\label{app:B}
In the following we give notes about the clusters we can define as outliers either in the 
$\tau_{HB}$-[Fe/H] plane or in the $\tau_{HB}$-age plane or both.
 
\textbf{NGC~6218}: outlier of $\tau_{HB}$-[Fe/H], while we cannot consider it an outlier in $\tau_{HB}$-age planes 
(it is located under the $2.0 \sigma$ levels). Both ground-based and space data are of good quality. It has been identified as a second parameter cluster.

\textbf{NGC~288}: outlier of $\tau_{HB}$-[Fe/H] and of the $\tau_{HB}$-age planes. Both ground-based and space 
data are of good quality. It has been identified as a second parameter cluster.

\textbf{NGC~6205}: outlier of $\tau_{HB}$-[Fe/H] and of the $\tau_{HB}$-age planes. Both ground-based and space 
data are of good quality. It has been identified as a second parameter cluster.

\textbf{NGC~6254}: outlier of $\tau_{HB}$-[Fe/H] and of the $\tau_{HB}$-age planes. Both ground-based and space 
data are of good quality. It has been identified as a second parameter cluster.

\textbf{NGC~6426}: outlier of $\tau_{HB}$-[Fe/H] and of $\tau_{HB}$-age planes. We have only space photometry. Among the second parameter clusters, it is the one
attaining the lowest value in $\tau_{HB}$.

\textbf{NGC~6752}: outlier of $\tau_{HB}$-[Fe/H] and of the $\tau_{HB}$-age planes. Both ground-based and space data 
are of good quality, but ground data do not reach the globular tidal radius. It has been identified as a second parameter cluster.

\textbf{NGC~6541}: outlier of $\tau_{HB}$-[Fe/H] and of the \citet{vand,leaman} $\tau_{HB}$-age plane. 
We have no age estimation from \citet{salarisweiss}. Both ground-based and space data are of good quality. It has been identified as a second parameter cluster.

\textbf{NGC~6681}: outlier of $\tau_{HB}$-[Fe/H] and of the $\tau_{HB}$-age planes. Both ground-based and space data 
are of good quality. It has been identified as a second parameter cluster.

\textbf{NGC~6362}: outlier of $\tau_{HB}$-age plane for age estimations from \citet{vand, leaman}.
Both ground-based and space data are of good quality.

\textbf{NGC~6535}: outlier of $\tau_{HB}$-age planes (above $2.5 \sigma$ limit in the \citet{vand,leaman} plane and
 $2.0 \sigma$ limit in the \citet{salarisweiss} one). Both ground-based and space data are of good quality.

\textbf{NGC~6652}: outlier of the \citet{salarisweiss} $\tau_{HB}$-age plane. Both ground-based and space data are of good quality.

\textbf{NGC~7089}: outlier of the \citet{vand,leaman} $\tau_{HB}$-age plane. We have no age estimation from 
\citet{salarisweiss}.
 Both ground-based and space data are of good quality.
 
\textbf{NGC~6171}: outlier in the $\tau_{HB}$-age planes (above the $2.5 \sigma$ limit in \citet{vand} and the $2.0 \sigma$ 
limit in \citet{salarisweiss}). Both ground-based and space data are of good quality.
\end{appendix}

\end{document}